\documentstyle[eqsecnum,prb,aps]{revtex}
\begin{document}
\draft
\columnsep-.375in
\twocolumn[
\title{Statistical Mechanics of Nonuniform Magnetization Reversal}
\author{Hans-Benjamin Braun\cite{address}}
\address{
Department of Physics, University of California San Diego, La
Jolla CA 92093-0319}
\date{ May 9, 1994 }
\maketitle
\begin{abstract}
\widetext
Thermally activated magnetization reversal in elongated
particles is studied within a model that allows for
spatially nonuniform magnetization configurations
along the particle. An external field antiparallel to the
existing magnetization is shown to give rise to an energy barrier
which represents a spatially localized deviation from the
initial  uniform magnetization configuration.
For sufficiently elongated particles
thermal fluctuations thus substantially lower the coercivity
compared to the previous theories by N\'eel and Brown
which assume a spatially uniform magnetization.
The magnetization reversal
rate is calculated using a functional Fokker-Planck description
of the stochastic magnetization dynamics.
Analytical results are obtained in the limits
of small fields and fields close to the anisotropy field.
In the former case the hard-axis anisotropy becomes
effectively strong and the magnetization  reversal rate
is shown to reduce to the nucleation rate of kink-antikink
pairs in the overdamped sine-Gordon model.
The present theory therefore includes the nucleation
theory of the double sine-Gordon chain as a special case.
\end{abstract}
]
\narrowtext
\section{introduction}
The magnetization in a uniformly magnetized sample is usually
stabilized by an easy-axis anisotropy of crystalline
or demagnetizing origin.
To reach a state of zero net magnetization one has to apply an external
field in the reversed direction, the so called coercive field.
In macroscopic samples of high purity such as
Yttrium Iron Garnet (YIG) \cite{Gyorgy}, this field  can be less than
$10^{-2}$ Oe. This low coercivity is
commonly attributed to the existence of residual
domains of reverse magnetization in the original
uniformly magnetized state. The measured coercivity is then
associated with the depinning and motion of the corresponding
domain walls.

This  situation is strikingly different for microscopic
single-domain particles where no such residual domains exist.
The coercivity can reach values of more than $1000$ Oe
since the state of reversed magnetization has first to be nucleated.
Consequently such particles exhibit an extremely high
long-term stability of the magnetization.  This fact renders them
suitable for information storage in recording media and as
constituents of rocks  they preserve the value of the
local magnetic field as the temperature has dropped
below the blocking temperature of the particle.
With decreasing sample size, however, the effect of
thermal fluctuations becomes increasingly important.
For particle sizes of a few nanometers and at room temperature
the magnetization fluctuates randomly over the anisotropy
barrier and a superparamagnetic state results with vanishing
average magnetic moment.

This paper  concentrates on particles whose size is above
the superparamagnetic limit but which are still small enough
that the coercivity is affected by thermal fluctuations.
The only ab-initio theory has been developed by
N\'eel \cite{neel}  and Brown\cite{brown} and it is based on
the assumption that the magnetization distribution is uniform
throughout the sample.   Consequently the energy barrier is
proportional to the volume and the Arrhenius factor leads to
an exponential suppression of thermal effects with the particle
volume.  This picture is indeed adequate for small particles of
approximately spherical shape.
However, for sufficiently elongated particles
a magnetization reversal via a rigid rotation of the magnetization
becomes energetically  unfavorable.  It will be
more advantageous to form a spatially localized excursion
from the metastable state since the
additional cost of exchange energy due to the spatial
nonuniformity is by far outweighed by the gain of anisotropy energy
by keeping the deviation localized.

It is the purpose of this paper to formulate an
{\it ab initio} theory of this effect and to show
that for sufficiently elongated particles a spatially
nonuniform barrier yields a much lower coercivity
than previous theories.
A short account of the results of the present
paper has already been given elsewhere \cite{prlnucl}.
We shall start from a classical one-dimensional
model energy density which takes into account the
exchange interaction between the magnetic moments
along the particle. In addition, the energy density
contains hard- and easy axis anisotropies
as well as the coupling to an external
field. The anisotropies may contain contributions of both,
shape and crystalline anisotropies.  The  barrier energy
is then shown to be independent of the hard-axis
anisotropy, and it is proportional to the domain
wall energy and the sample cross sectional area.
Consequently, the barrier energy is
independent of the particle length for
sufficiently elongated particles.

In order to induce magnetization reversal, thermal fluctuations
have to form a ``nucleus" of critical size with the property that
smaller deformations fall back to the metastable state
whereas larger deformations grow with energy gain
until the magnetization is reversed.  Therefore the nucleus
represents an unstable structure with exactly one unstable mode.
An analytical expression for this structure has been
obtained \cite{prlnucl,broz} and  the spin wave excitations of
the nucleus have been investigated in a previous
paper \cite{braun1} (henceforth referred to as I).  For
external fields close to the anisotropy field which renders
an individual magnetic moment unstable,  the nucleus
represents only a small  deviation from the
metastable state. For small fields, the nucleus consists
of two well separated domain walls enclosing an already
reversed domain.

The present approach relies on methods that have been used
in the description \cite{Langer} of the dynamics of first order
phase transitions. This method has been applied  for the
description of the decay of a supercurrent  \cite{Halp} in a
thin wire  or the propagation \cite{Pokr,Butt} of  dislocations.
In contrast to these applications we consider here the
regime of moderate damping since damping in magnetic systems is
very small. The rate is shown to be the product of a prefactor
depending on the external field and
temperature $T$,  and the Arrhenius factor
$\exp \{-{\cal A E}_s /k_B T\}$ which involves
the barrier energy ${\cal A} {\cal E}_s$ with ${\cal A}$
the sample cross sectional area  ($k_B$ is the Boltzmann constant).

For the evaluation of the prefactor
we shall employ two different approaches.
One is the Jacobi method which relies on  the explicit  knowledge of
a zero-energy (Goldstone)  mode. Therefore it can  only be applied for
easy-plane fluctuations  but it cannot be used  for out of  easy-plane
excitations due to the existence of a mass gap.  The second
method makes use of the scattering phase shifts of  spin
waves around the  nucleus. This latter approach reveals some
considerable subtleties which do not seem to have
been  noted previously.  First, for even-parity
wavefunctions the 1D version \cite{barton} of  Levinson's theorem
alters the usual  expression for the density of states.
Second, the number of bound states of the fluctuation operators
is not conserved under small and large nucleus approximations.
This fact raises doubts about the commonly employed approach
of performing functional integrals of the free energy
after already having performed the limits of small or large nuclei.
By a careful investigation it is shown that these two
subtleties conspire in such a way that this interchange of limits is
indeed legitimate.

There are only sparse treatments of magnetization reversal
in the literature. However, this field is closely related
to macroscopic quantum tunneling
of the (sublattice-) magnetization  in  small (anti-)ferromagnetic
grains, a subject that has attracted much interest
recently.  It thus appears to be useful to relate some important
papers that contributed to the development of these fields.

Early work on nucleation theory
culminated in the celebrated  paper \cite{Kramers}
by Kramers who calculated the escape rate due to
thermal activation out of a metastable state
in the limits of low as well as moderate to large friction.
He showed that the rate is given by a prefactor times the
Arrhenius term. Despite the fact that his work was
restricted to one degree of freedom,
his method of the evaluation of the prefactor turned out to be
so powerful that its spirit still underlies much more
complex applications. An extension to an
arbitrary number of degrees of freedoms in the large friction
limit  was due to Brinkman, Landauer-Swanson  and
Langer \cite{Brink}. The case of moderate friction
has been considered by Langer \cite{Langer}
who also pointed out  that  the nucleation rate
may be interpreted as the analytic continuation
of the partition function.  This idea is closely related to
the subsequently developed instanton
concept \cite{calcol} in Euclidean quantum field theories.
Kramers's theory and its extensions have recently
been reviewed by H\"anggi, Talkner and Borkovec \cite{hanggi}.

The first application of Kramers's theory to magnetic systems
has been made by  Brown \cite{brown} who
investigated  thermally activated uniform magnetization
reversal in small ferromagnetic particles to explain
superparamagnetism.
He set up the  Fokker-Planck equation for the stochastic
dynamics of the magnetization and thus related N\'eel's
earlier considerations \cite{neel} on reversal rates
with the general framework of statistical mechanics.
For axial symmetry of the anisotropy he
obtained nucleation rates from the lowest nonzero  eigenvalue
of the Fokker-Planck equation in the limit of low barriers.
For a high barrier he used Kramers's  procedure to
evaluate the rate constant. Later the lowest positive
eigenvalue was investigated  numerically for all intermediate
values between low and large barriers by Aharoni \cite{aharoni}.
Eisenstein and Aharoni \cite{eisah} investigated the
competition of the  uniform mode and the nonuniform
curling mode as possible candidates of critical nuclei
for different particle radii.  However,  the nucleation rates
via the nonuniform mode were  calculated using
Brown's theory \cite{brown} for spatially uniform nucleation.

Subsequently the issue of magnetization reversal
rates was not addressed for many years.  A renewal
of interest  then arose from a quantum mechanical
point of view.  Based on path integral \cite{enzschilling}
and WKB \cite{Hemm} techniques, first investigations
showed that a single spin in an anisotropic field
behaves similar to a particle with inertia
and tunnels between different anisotropy
minima.  It has then been suggested \cite {ChudGunti} that
in small ferromagnetic particles
macroscopic quantum tunneling might occur.
The effect of dissipation due to magnetoelastic
coupling has been discussed by Garg and Kim \cite{GaKi}.
In the context of recent experiments \cite{Awsch} these
approaches have been reexamined and it has been
predicted \cite{Loss} that quantum tunneling is
suppressed for half integer spins as a consequence
of a previously neglected  Wess-Zumino term
in the quantum spin action and the destructive interference of
instanton and antiinstanton paths.

While all these approaches dealt with tunneling via
spatially uniform structures, tunneling via spatially
nonuniform (bubble) structures in two dimensions
was investigated in the limit of external magnetic fields
close to the anisotropy field \cite{Chudii}  and for
very small fields in the thin wall approximation \cite{CaldFur}.
In the latter case the nucleating structure is a
large cylindrical domain of reversed magnetization
delimited by a Bloch wall.
Various aspects of quantum tunneling with emphasis
on tunneling of Bloch walls have recently been
reviewed by Stamp, Chudnovsky and Barbara \cite{Stamp}.

Surprisingly, the conceptually much simpler classical problem
of thermal nucleation  remained untouched until recently.
Klik and Gunther \cite{KlGu} calculated the nucleation rate
for nucleation via uniform structures for cubic symmetry.
In contrast to earlier investigations they also calculated
nucleation rates for a weakly damped system.
Nuclei of curling symmetry in an infinite cylinder \cite{Broz2}
and a nucleation center of spherical symmetry \cite{AhaBalt}
have been investigated recently.

{}From this review, there emerges clearly the need of an
{\it ab initio} theory for magnetization reversal rates
via spatially nonuniform structures.
The present work is organized as follows:

In section II  some results of  paper I  are reviewed
which are relevant to the present work.
In section III  a functional Fokker-Planck equation
is constructed which describes the stochastic magnetization
dynamics  near the nucleus and the corresponding nucleation
rate is derived.
It is shown  that the result has the same general
structure as that of
Ref.  \onlinecite{Langer}(b).
The prefactor separates into  a
term describing the dynamical decay of the nucleus
and in a term arising from the Gaussian fluctuations
around the nucleus. The unstable mode enters in such a
way as if it represented a stable mode. The details of
the calculations are presented in an appendix.
The explicit evaluation of the prefactors in various limits
is then the subject of the remaining sections.
In section IV we evaluate the statistical part of
the prefactor analytically in the limit of small and large nuclei
as well as for large and small values of the hard-axis
anisotropy.
For small nuclei and if the hard-axis anisotropy
is much larger than the easy-axis
anisotropy, the out of easy-plane fluctuations
do not contribute at all.
In section V the nucleation rate is evaluated in the
overdamped limit.
The rate in the  moderately damped regime and the
decay frequency of the nucleus are investigated in section  VI.
In section  VII the results of the previous sections
are used to calculate the creation rate of
kink-antikink pairs in the  double sine-Gordon
system. It is shown that this rate
reproduces the magnetization reversal  rate in the
limit of large hard-axis anisotropy or external fields
close to the (easy-axis) anisotropy field.
In section VIII  experimental implications
are discussed. For a particle of $100\AA$ diameter
and an aspect ratio $15:1$,  the present theory is
shown  to yield a coercivity reduction from the
anisotropy field that is twice as large as that
of the N\'eel-Brown theory.
Finally  the applicability range of the
present theory is discussed since it is known that  in the
underdamped  limit the rate is governed by a
diffusion in energy rather than in configuration space.

\section{model, nucleus and fluctuations}

In this section we present the model and review some
important results of  paper I. The ferromagnet is
described within a classical continuum model,
the magnetization being represented by a
vector ${\bf  M}$ of constant magnitude $M_0$.
We focus on an effectively one-dimensional situation
where the magnetization only depends on  one coordinate,
i.e. ${\bf  M}={\bf  M}(x,t)$. The energy per unit area
is given by
\begin{eqnarray}
 {\cal E}=\int_{-L/2}^{L/2} dx   \left\{ {A\over M_0^2}
         \left[ (\partial_x M_x)^2+
               (\partial_x M_y)^2+(\partial_x M_z)^2\right]
          \right.  \nonumber\\
          \left.      -{K_e\over M_0^2} M_x^2
                  +{K_h\over M_0^2}   M_z^2-
          H_{\rm ext} M_x  \right\},
\label{eb1}
\end{eqnarray}
where $\partial_x\equiv\partial/\partial x$ and
$L$ is the finite sample length in $x$-direction.
Ultimately, we shall be interested in the limit $L\to\infty$.
The first term in (\ref{eb1}) is the classical counterpart of
exchange energy and $A$ is an exchange constant.
The second term defines an easy-axis along the $x$-direction.
The third term is a hard-axis anisotropy which
favors the magnetization to lie in the $xy$-plane.
$K_e>0$ and $K_h>0$ are easy- and hard-axis anisotropy
constants respectively.  The degeneracy
between the two  anisotropy minima  along the
$x$-axis is lifted by an external magnetic field
$H_{\rm ext}$ along the positive  $x$-axis.
The energy (\ref{eb1})  describes  magnetization
configurations in an elongated particle of diameter
smaller\cite{onedim} or comparable to the minimal
length scale in the system $\sqrt{A/K_{\rm max}}$,
where $K_{\rm max}$ is the larger of the anisotropy
constants $K_e, K_h$.

Note that the  energy  (\ref{eb1})  can be used to
describe three distinct anisotropy configurations in
elongated particles.
The first, most common case is an easy-axis along the
particle axis which may be caused by both demagnetizing
(shape) and crystalline anisotropy (cf. Fig \ref{fig0} a).
E.g. for an infinite cylinder with an easy-axis along
the sample one has $K_e=\pi M_0^2 +
K_{e,{\rm cryst}}$ where  the first term is due to the
shape anisotropy and the second term
 due to crystalline anisotropy.
The  hard-axis anisotropy  may
arise either from an additional crystalline
easy-axis that is misaligned
with the particle axis or from an elliptic sample cross section.
The second case is an elongated particle of a material
of high crystalline anisotropy ($K_{e,{\rm cryst}}>
2\pi M_0^2$) with both easy- and hard-axis perpendicular
to the long-axis of  the sample  (cf. Fig \ref{fig0}b).
The third case refers to a thin slab with an easy-axis anisotropy
in the film plane  (cf. Fig \ref{fig0} c).

In the following we focus on a situation as in Fig. 1a).
The results for configurations shown in  Figs. 1 b),c)
are simply obtained by substituting $y$ and $z$
for the $x$-dependence of the magnetization.
The components in internal (spin) space remain unchanged and
the spherical coordinates are always defined in the same way
with respect to the coordinate axes.

The dissipative dynamics of the magnetization is assumed to
obey the Landau-Lifshitz-Gilbert equations (see e.g.
Ref. \onlinecite{Enzetal}):
\begin{equation}
{\partial_t {\bf M}} = -\gamma {\bf M}\times {\bf H}_{\rm eff}
+{\alpha\over M_0}{\bf M}\times \partial_t {\bf M}.
\label{LLG0}
\end{equation}
where $\gamma>0$ denotes the gyromagnetic ratio,
$\alpha>0$ is the dimensionless damping constant  and
$\partial_t=\partial/\partial t$.
The first term on the r.h.s. of (\ref{LLG0}) describes
the precession of the magnetization in
the effective magnetic field ${\bf H}_{\rm eff} =
 - \delta {\cal E}/ \delta {\bf M}$ ($\delta/\delta {\bf M}$
denotes a functional derivative).
The second term in (\ref{LLG0}) is a viscous damping term
and accounts for the relaxation of the magnetization into
the direction of the effective magnetic field.
This term is phenomenological in nature. It describes
damping processes which conserve the magnitude of the
magnetization at every space point.
It is convenient to rewrite (\ref{LLG0}) as follows
\begin{equation}
(1+\alpha^2){\partial_t{\bf M}} = -\gamma {\bf M}\times
{\bf H}_{\rm eff} -{\alpha\gamma\over
M_0}{\bf M}\times[{\bf M}\times{\bf H}_{\rm eff}].
\label{LLG}
\end{equation}
This equation is obtained by evaluating the cross product of
${\bf M}$ with (\ref{LLG0}).  Eq. (\ref{LLG}) is similar to
the damping term originally proposed by Landau and Lifshitz.
However,  Eq. (\ref{LLG}) contains the damping parameter
$\alpha$ such that the motion is slowed
down for large $\alpha$ while the original equation
of Landau-Lifshitz  exhibits an unphysical acceleration
of the motion for large damping parameters.

Since (\ref{LLG}) conserves the magnitude of the
magnetization it is appropriate to
introduce spherical coordinates according to $\;{\bf M}/M_0$
$= (\sin\theta\cos\phi$, $\sin\theta\sin\phi,\cos\theta)\;$.
We use dimensionless units defined by
\begin{eqnarray}
\left[x\right]&=&\left[y\right]=\left[z\right]=
\sqrt{{A\over K_e}},\nonumber\\
\left[t\right]&=&(1+\alpha^2){M_0\over 2\gamma K_e},
\nonumber\\
\left[{\cal E}\right]&=&2\sqrt{AK_e}.
\label{units}
\end{eqnarray}
$\sqrt{A/K_e}$ is the Bloch wall width,
$2\gamma K_e/M_0$  is the precession
frequency in the anisotropy field. To simplify notation,
an additional factor $1+\alpha^2$ has been absorbed
in the time scale.  $2\sqrt{AK_e}$ is half the energy
of a  static $\pi$-Bloch wall.  In dimensionless units and
spherical coordinates the energy per area (\ref{eb1}) takes the
following form
\FL
\begin{eqnarray}
{\cal E}=\int_{-L/2}^{L/2}\!\! dx \left\{  {1\over 2}
     \left[(\partial_x\theta)^2+\sin^2\theta (\partial_x \phi)^2
     \right]
      \right.\nonumber\\
 \left.-{1\over2} [\sin^2\theta\cos^2\phi -1]
+{Q^{-1}\over 2} \cos^2\theta-h[\sin\theta\cos\phi +1] \right\},
\label{e}
\end{eqnarray}

In (\ref{e}),  the dimensionless anisotropy ratio
\begin{equation}
Q={K_e\over K_h},
\label{Q}
\end{equation}
and the reduced external field
\begin{equation}
h={H_{\rm ext}M_0\over 2 K_e},
\label{h}
\end{equation}
have been introduced. Using (\ref{units}) and  spherical coordinates,
the equations of motion (\ref{LLG}) take the following
form (see also the appendix of Ref. \onlinecite{Enzetal})
\begin{eqnarray}
\sin\theta\, \partial_t\phi& =&
{\delta {\cal E}\over \delta \theta} -\alpha {1\over \sin\theta}
{\delta{\cal E}\over\delta\phi} \nonumber\\
 \partial_t \theta&=&-{1\over \sin\theta}{\delta{\cal E}
\over\delta\phi}-\alpha{\delta {\cal E}\over\delta \theta}
\label{sLLG}
\end{eqnarray}
The first terms on the r.h.s. describe the precession in the effective
magnetic field, whereas the terms proportional  to $\alpha$ are
damping terms.

Spatially uniform static solutions of  (\ref{LLG})
lie in the easy plane and are given
by $(\phi_0,\theta_0)=(0,\pi/2)$
and $(\phi_m,\theta_m)=(\pi,\pi/2)$,
the latter being stable only for $h<1$. The state
$(\phi_0,\theta_0)$ is completely aligned with the external
field and thus represents the state of lowest energy.
The configuration $(\theta_m,\phi_m)$ is oriented
antiparallel to the  external field and its energy per volume
exceeds that of the ground state $(\phi_0,\theta_0)$ by $2h$.
Therefore $(\theta_m,\phi_m)$ is a metastable state for $h<1$.

At finite temperatures, the magnetization exhibits
fluctuations around the metastable state until it
eventually overcomes a barrier for magnetization reversal.
For large sample lengths $L$, a magnetization reversal
via a uniform rotation of the magnetization is highly unlikely
since it would require an energy proportional to $L$.
Instead, the system will establish magnetization reversal
by forming a spatially localized  deviation from the
metastable state. There is a well defined  ``nucleus"  of
critical size with the property that
deformations of smaller size fall back to the metastable state,
whereas larger deformations grow with energy gain until
the whole system is in the ground state parallel to the
external field. In paper I it has been
shown that the magnetization configuration (cf. (I,3.9))
defined by
\begin{equation}
\tan{\phi_s^\pm\over 2}=\pm{\cosh{x-x_0\over\delta}\over
\sinh R},
\qquad\theta_s=\pi/2,
\label{nucl}
\end{equation}
with
\begin{equation}
\mathop{\rm sech}\nolimits^2 R=h,\qquad \delta=\coth R,
\label{rdef}
\end{equation}
exhibits exactly one unstable mode and thus represents such
a nucleus of critical size.
Eq. (\ref{nucl}) is in principle only valid for a sample
of infinite length $L$ but it is an excellent approximation
for a sample of finite length if $L>2\pi \sqrt{A/K_e}$.
For typographical simplicity the subscript $s$ of $\delta$
and $R$ has been dropped in contrast to paper $I$.
$x_0$ denotes the arbitrary position of the nucleus along
the particle. This degeneracy with respect to
translations gives rise to zero energy (Goldstone) mode.
In the following we shall  put $x_0=0$.
The structure (\ref{nucl}) can also be viewed
as a superposition of two $\pi$-Bloch walls centered at
$ x=\pm R /\delta +x_0$ with opposite  relative sense of twist.
For $R$ small, (\ref{nucl}) represents only
a small deviation from the metastable state, whereas for large
$R$ it represents a large  domain of size $2R\delta$
delimited by an untwisted pair of domain walls (cf. Fig. \ref{fig1}).
In the following we shall restrict ourselves to $\phi_s^+$ only
and we shall drop the  superscript. The existence of two
equivalent saddle points will result in a factor of two in the final
expression of the nucleation rate.

Out of easy-plane fluctuations $p$ and azimuthal
fluctuations $\varphi$ around $\phi_s$ are introduced as follows
\begin{eqnarray}
\phi(x,t)&=&\phi_s(x)+\varphi (x,t),\nonumber \\
\theta(x,t)&=&\pi/2 -p(x,t).
\label{nucllin}
\end{eqnarray}
Inserting (\ref{nucllin}) into the energy (\ref{e})
we obtain up to $2^{nd}$ order in $\varphi$ and $p$
\FL
\begin{eqnarray}
{\cal E}^{(2)}_s={\cal E}_s + {1\over 2} \int_{-L/2}^{L/2}
\!dx\;\varphi\;{\cal H}^{s\varphi}\;\varphi
+ {1\over 2} \int_{-L/2}^{L/2} \!dx\; p\;{\cal H}^{s p}\;p,
\label{e2nucl}
\end{eqnarray}
where
\begin{eqnarray}
{\cal H}^{s\varphi}&=&-{d^2\over dx^2}+\delta^{-2}V_-
\left({x\over\delta},R\right),\label{hfop}
\\
{\cal H}^{s p}&=&-{d^2\over dx^2}+\delta^{-2}V_+
\left({x\over\delta},R\right)+Q^{-1}.
\label{hpop}
\end{eqnarray}
The energy per unit area of the nucleus (\ref{nucl}) is given by
\begin{equation}
{\cal E}_s=4\tanh R - 4R \mathop{\rm sech}\nolimits ^2 R.
\label{nuclen}
\end{equation}
The characteristic width $\delta$ is given
by  (\ref{rdef}), and the potentials $V_{\pm}$  can
be inferred from (I,4.13)
\begin{eqnarray}
V_\pm (\xi,R)&=&1-2\mathop{\rm sech}\nolimits^2(\xi+R)-
2\mathop{\rm sech}\nolimits^2(\xi-R)\nonumber\\
&\pm& 2\mathop{\rm sech}\nolimits(\xi+R)
\mathop{\rm sech}\nolimits(\xi-R).
\label{V}
\end{eqnarray}
The eigenvalue problems of (\ref{hfop}) and (\ref{hpop}) are
written as follows
\begin{eqnarray}
{\cal H}^{s\varphi}\chi^{s\varphi}_\nu (x,R) &=&
E^{s\varphi}_\nu(R)\chi^{s\varphi}_\nu(x,R),
\label{Hsfev}\\
{\cal H}^{s p}\chi^{sp}_\nu (x,R) &=&E^{sp}_\nu(R)
\chi^{sp}_\nu(x,R),
\label{Hpfev}
\end{eqnarray}
where $\nu$ denotes both, bound states and scattering states.
{}From I, two solutions of the eigenvalue problems (\ref{Hsfev})
are known:
the ground state of ${\cal H}^{s p}$
\begin{equation}
\chi^{sp}_0\propto\delta^{-1}\left\{\mathop{\rm sech}
\nolimits({x\over \delta}+R)
+\mathop{\rm sech}\nolimits({x\over\delta}-R)\right\},\;\;
E_0^{sp}=Q^{-1},
\label{cp1}
\end{equation}
and the first excited state
of ${\cal H}^{s\varphi}$
\begin{equation}
\chi^{s\varphi}_1\propto\delta^{-1}\left\{\mathop{\rm sech}
\nolimits({x\over \delta}+R)
-\mathop{\rm sech}\nolimits({x\over\delta}-R)\right\},\;\;
 E_1^{s\varphi}=0.
\label{cf1}
\end{equation}
Since $\chi^{s\varphi}_1$ is antisymmetric, there is exactly
one unstable mode of negative energy in $\varphi$ while
all fluctuations in $p$-direction have positive energy
since $Q^{-1}>0$. Therefore the untwisted domain wall pair
represents a saddle point of the energy with exactly one
unstable direction.

The operators characterizing the modes around the metastable state
$(\phi_m,\theta_m)$ are obtained in an analogous way. Inserting
\begin{eqnarray}
\phi(x,t)&=&\pi+\varphi (x,t),\nonumber\\
\theta(x,t)&=&\pi/2 -p(x,t),
\label{homlin}
\end{eqnarray}
into (\ref{e}) we have
\FL
\begin{equation}
{\cal E}^{(2)}_m= {1\over 2} \int_{-L/2}^{L/2} \!dx\;\varphi\;
      {\cal H}^{m \varphi}\;\varphi
                 + {1\over 2} \int_{-L/2}^{L/2} \!dx\; p
     \;{\cal H}^{m p}\;p,
\label{e2meta}
\end{equation}
where  the operators
\begin{eqnarray}
{\cal H}^{m\varphi}&=&-{d^2\over dx^2}+\delta^{-2},\nonumber\\
{\cal H}^{m p}&=&-{d^2\over dx^2}+\delta^{-2}+Q^{-1},
\label{hmop}
\end{eqnarray}
reflect the spatial uniformity of the metastable state.
In order to calculate nucleation rates by thermal activation,
we have to examine the stochastic dynamics around the nucleus.

\section{stochastic motion and nucleation rate}

The dissipative dynamics of the magnetization is
governed by the equations of motion (\ref{sLLG}).
To investigate the dynamics near the nucleus,
we insert (\ref{nucllin}) and (\ref{e2nucl}) into
(\ref{sLLG}) and obtain the linearized equations of motion
\begin{eqnarray}
\partial_t\varphi&=-{\cal H}^{s p} p-\alpha{\cal H}^{s\varphi} \varphi,
\nonumber\\
\partial_tp&={\cal H}^{s\varphi} \varphi-\alpha{\cal H}^{s p} p.
\label{linLLG}
\end{eqnarray}
Notice the unusual occurrence of
damping terms proportional to $\alpha$ in the equations
of motion for both $\varphi$ and $p$.
The reversible part of
(\ref{linLLG}) is of Hamiltonian structure. This
is due to the fact that the $z$-component of angular momentum
$\cos\theta$ and the azimuthal angle $\phi$ are canonically
conjugate variables. However, the signs in Eq. (\ref{linLLG})
 are reversed compared
to a usual canonical theory since we are dealing with the
magnetization rather than the angular momentum.

The statically unstable mode has a dynamical
counterpart $(\varphi_+(x,t),p_+(x,t))
\propto e^{\lambda_+ t} (\varphi_+(x),p_+(x))$ with
$\lambda_+>0$  which inserted into (\ref{linLLG})  obeys
the coupled eigenvalue problem:
\begin{eqnarray}
\lambda_+ \varphi_+&=&-{\cal H}^{s p} p_+ -\alpha {\cal H}^{s\varphi}
\varphi_+,\nonumber\\
\lambda_+ p_+&=&{\cal H}^{s\varphi}\varphi_+-\alpha{\cal H}^{s p} p_+,
\label{lambda}
\end{eqnarray}
with the boundary conditions
$\varphi_+(\pm {L\over 2})=p_+(\pm {L\over 2})$
$=\varphi_+'(\pm{L\over 2})=p_+'(\pm{L\over 2 })=0$.
The linearized equations  (\ref{linLLG})
of motion can also be cast in a compact form
\begin{equation}
\partial_t \psi_i (x,t)=-\sum_j M_{ij} {\cal H}_j\psi_j (x,t).
\label{psilin}
\end{equation}
In (\ref{psilin}) we have introduced
\begin{equation}
\psi(x,t)\equiv (\varphi(x,t),p(x,t))
\label{psidef}
\end{equation}
and  the dynamic matrix
\begin{equation}
M=\left(\begin{array}{cc}
\alpha &1 \\
 -1&\alpha
\end{array}\right).
\label{M}
\end{equation}
which  is the sum of a symplectic matrix describing the
reversible part of the dynamics and a diagonal positive
definite dissipative matrix.
For the operators in (\ref{psilin}) we have used the notation
$({\cal H}_1,{\cal H}_2)\equiv ({\cal H}^{s\varphi},
{\cal H}^{s p}).$

Eqns (\ref{psilin}), and equivalently (\ref{linLLG}), describe
the deterministic motion of  the system in the vicinity of the saddle point.
However, they are not consistent with
the fluctuation dissipation theorem since they lack
the stochastic forces resulting from the coupling
to the heat bath. Without these stochastic forces, the magnetization
would never be driven away from the initial metastable state.
Stochastic forces can be added to the r.h.s. of  (\ref{linLLG}) or
(\ref{psilin}) to yield the Langevin equation
\begin{equation}
\partial_t \psi_i (x,t)=-\sum_j M_{ij} {\cal H}_j\psi_j (x,t)
+ \zeta_i(x,t)
\label{langevin}
\end{equation}
where $\zeta_i$ is Gaussian white noise
with $\langle\zeta_i\rangle=0$ and
\begin{equation}
\langle \zeta_i(x,t) \zeta_j(x',t')  \rangle=
{ 2\alpha\over\beta {\cal A} }\delta_{ij}\delta(x-x') \delta(t-t')
\label{corr}
\end{equation}
where $\langle \dots\rangle$  denotes the average
with respect to  the Gaussian noise distribution
$\exp \{  -\beta {\cal A}/( 4\alpha) \int dt dx \sum_i \xi_i^2 \}$
and $\beta = 1/k_B T$. The dynamics  of the probability distribution
functional $\varrho[\psi(x)]=$ $\langle \prod_{i,x} \delta (\psi_i(x,t)-
\psi_i(x) )\rangle$ with $\psi_i(x,t)$ a solution of (\ref{langevin}) is
governed \cite{zinn} by the Fokker-Planck equation
\begin{equation}
\partial_t \varrho [\psi(x)]=-\int dx
\sum_i {\delta J_i\over\delta\psi_i(x)}.
\label{FP}
\end{equation}
The probability current is given by
\begin{equation}
J_i= -\sum_j  M_{ij} \left[{\cal H}_j \psi_j(x) +{1\over\beta{\cal A}}
{\delta \over\delta\psi_j (x)}\right]\varrho[\psi(x)],
\label{current}
\end{equation}
In (\ref{current})  we have  exploited the antisymmetry
of the off-diagonal part of $M$.
Note that the current is only defined up to a
divergenceless term. If we demand in addition that the equilibrium density
has vanishing current, the representation (\ref{current}) is unique.
Equations (\ref{FP}) and (\ref{current})  have the following properties:

i) The equations of motion for the thermal expectation values
$\left<\varphi\right>$,$\left<p\right>$ are identical to the expectation
 value of (\ref{psilin}).
[ Expectation values are defined by
$\left<\varphi\right>=\int{\cal D}\varphi{\cal D} p\;\varphi\,\varrho$,
where $\int {\cal D}\varphi$ denotes functional
integration.]

ii) The equilibrium density  near the saddle point
\begin{equation}
\varrho_{\rm eq}=Z^{-1}\exp\{-\beta{\cal A}{\cal E}_s^{(2)}\},
\label{req}
\end{equation}
with ${\cal E}_s^{(2)}$ as in (\ref{e2nucl}), is a stationary solution
of (\ref{FP}) with vanishing current. $Z$ is a normalization constant arising
from the condition that $\int{\cal D} \varphi{\cal D} p \,
\varrho_{\rm eq} =1$ in the vicinity of the metastable state.
Since $\varrho_{\rm eq}$ is sharply peaked around the
metastable state, a Gaussian approximation
may be used for the evaluation of $Z$.
Note that the properties i), ii) also allow for a direct construction of the
Fokker-Planck equation without making use of the Langevin equation.

To calculate  the nucleation rate, we have to construct
a stationary nonequilibrium probability density.
To maintain a constant probability flux over the saddle
point we impose the boundary conditions  $\varrho
\simeq \varrho_{\rm eq}$ near the metastable state
and $\varrho\simeq 0$ beyond the saddle point.
Note that the realization of equilibrium at the metastable state
requires a barrier energy which should be large compared to
thermal energies. As a criterion we may use
$\beta {\cal A} {\cal E}_s \raise1pt\hbox{$>$}
\lower3pt\hbox{\llap{$\sim$}}5$. Since the prefactor
is roughly of the order of the precession frequency
$ 2\gamma K_e/M_0 \simeq 10^{10} s^{-1}$,
this inequality is satisfied even for very large switching rates and
thus does not represent a restriction. The total rate is then obtained
as the probability flux integrated across a surface transversal
to the unstable mode.  The derivation is similar to that of Langer
\cite{Langer} and  is presented in detail in the appendix.
The switching probability
per unit time of a particle with magnetization prepared in the
metastable state into the stable state is then given by
\begin{equation}
\Gamma=\Omega e^{-\beta{\cal A}{\cal E}_s},
\label{rate}
\end{equation}
where ${\cal A}$ is the cross
sectional  area of the sample and ${\cal E}_s$ is the energy
(\ref{nuclen}) per area of the nucleus.
The prefactor is given as follows
\FL
\begin{equation}
\Omega=\lambda_+{\cal L} \sqrt{{\beta{\cal A}\over 2\pi^3}}
\sqrt{{\det {\cal H}^{m\varphi}\over
\det' |{\cal H}^{s\varphi}|}}
\sqrt{{\det {\cal H}^{m p}\over
\det {\cal H}^{s p}}}.
\label{pref}
\end{equation}
In (\ref{pref}) an explicit factor of 2 has been included, since
the metastable state $\phi=\pi,\theta=\pi/2$ may decay
via either one of the two equivalent saddle points $\phi_\pm$.
The first factor on the r.h.s. of (\ref{pref}) is the escape
frequency of the unstable mode as defined by (\ref{lambda}).
This is the only term in (\ref{pref}) in which dynamical details
of the system enter. The second  factor arises from the integration
over the zero frequency (Goldstone) mode and is defined by
${\cal L}=\sqrt{{\cal E}_s} L$,
where $L$ is the system length and ${\cal E}_s$ is the
energy per unit area (\ref{nuclen}). The third factor is also
due to the Goldstone mode and determines the temperature
dependence of the prefactor. The remaining factors basically
arise from the functional integration of the partition function within
Gaussian approximation (\ref{e2nucl}) and (\ref{e2meta}).
The determinants are defined as the products of eigenvalues
\begin{equation}
{\det {\cal H}^{m\varphi}\over
\det' |{\cal H}^{s\varphi}|}
={\prod_{k} E^{m\varphi}_{k}
\over  |E_0^{s\varphi}| E_2^{s\varphi} \prod_{k'}E^{s\varphi}_{k'}},
\label{fdet}
\end{equation}
where $k$ denotes scattering states.
The prime on $\det$ denotes omission of the zero energy $E_1^{s\varphi}$.
However, note that the unstable mode $E_0^{s\varphi}<0$
enters (\ref{fdet}) as if it were a stable one.
The determinant of out of easy-plane fluctuations is defined as
\begin{equation}
{\det {\cal H}^{m p}\over
\det {\cal H}^{sp}}
={\prod_{k} E^{mp}_{k}
\over E_0^{sp} E_1^{sp} \prod_{k'}E^{sp}_{k'}}.
\label{pdet}
\end{equation}
The bound state energies $E^{s\varphi}_0$ and $E^{sp}_1$ are given by
(\ref{cp1}) and (\ref{cf1}). In  paper I we obtained $E^{s\varphi}_2\simeq
\delta^{-2}$ (I,6.13) and   $E^{s\varphi}_0$ and
$E^{sp}_1$ have been evaluated numerically for arbitrary $R$.

Therefore we are left with the task of evaluating the
products of the continuum eigenvalues.
The continuum eigenvalues of ${\cal H}^{s\varphi}$ and ${\cal H}^{s p}$
coincide with those of ${\cal H}^{m\varphi}$ and ${\cal H}^{mp}$, respectively,
and are given by
\begin{eqnarray}
E^{m\varphi}_k& =&E^{s\varphi}_k =\delta^{-2}+k^2,\nonumber\\
E^{m p}_k &=&E^{sp}_k =Q^{-1}+\delta^{-2}+k^2.
\label{eksp}
\end{eqnarray}
Note that these equalities do not imply a cancellation of numerators
and denominators in (\ref{fdet}) and (\ref{pdet}), since the
allowed $k$ -values are different and fixed by the
bondary conditions which we choose as periodic.

The next sections are devoted to the explicit evaluation
of the infinite products (\ref{fdet}), (\ref{pdet}) and
the calculation of the escape frequency $\lambda_+$.

\section{Evaluation of the statistical prefactor}

In the following we describe two methods for the evaluation
of the statistical prefactors (\ref{fdet}), (\ref{pdet}).

The first method (see e.g. Ref. \onlinecite{Raj}) is based on the
knowledge of scattering phase shifts
of the continuum eigenfunctions as well as the bound state
energies of the operators ${\cal H}^{s\varphi}$, ${\cal H}^{s p}$.
In strictly one dimensional  problems one has to
distinguish between the scattering phase shifts of odd and even
parity  wavefunctions.   This is in sharp contrast to the familiar
situation of a  3D $s$-wave scattering problem
where the scaled wavefunction indeed obeys a 1D Schr\"odinger
equation but is required to vanish at the origin.
Surprisingly, this issue  has been ignored until
recently \cite{barton,newton},  possibly also due to the fact
that the most widely used 1D potentials
belong to the rather special class of reflectionless
potentials  for which scattering phase shifts of even and odd
wavefunctions coincide.

So far, the present method for the  evaluation of the prefactor
has indeed been used only \cite{Raj} in the case of reflectionless
potentials where the scattering phase shifts of even and odd
wavefunctions coincide.  Here, however, the potentials in
${\cal H}^{s\varphi}$ and ${\cal H}^{s p}$ only become
reflectionless in the limits $R \to 0$ and $R\to \infty$.
For intermediate values of $R$, the corresponding potentials
are not reflectionless and scattering phase shifts of even and
odd parity wavefunctions have to be distinguished.

There is another surprising feature of these operators.
The number of bound states of the operators which arise in the
limits $R\to 0$ and $R\to \infty$ differs from those for
finite values of $R$.  This  casts some doubts on the usefulness of
such operators for an approximation  of the products (\ref{fdet}),
(\ref{pdet}). However, by a careful analysis using the explicit
form of Levinson's theorem in 1D,  we show  that the
exact fluctuation determinants converge to  those evaluated by
means of the limiting operators.

The second method uses the explicit knowledge of the zero
mode for the evaluation of the fluctuation determinants.
This allows an exact determination of the statistical prefactor
for the $\varphi$-fluctuations. For $p$-fluctuations an analytical
treatment is only possible in the limit $Q^{-1}\to 0$.

These results can then be combined to obtain analytical
expressions for the total statistical prefactor in the limit
of small and large nuclei as well as in the limits $Q^{-1}\to 0$
and $Q^{-1}\to\infty$.  In the limit $Q^{-1} \delta^2 \to \infty$
corresponding to either small nuclei or large hard-axis anisotropy,
the out of easy-plane fluctuations are suppressed and  do not
contribute at all.  While the latter result is to be expected from
the fact that out of easy-plane fluctuations are suppressed due to
their mass $Q^{-1}$,  the former result
is somewhat surprising. It is related to the divergence of the
characteristic length scale for $R\to 0$ which renders even  a
small hard-axis anisotropy effectively large.
In both limits the system may be described by an
effective model discarding the out of easy-plane degree
of freedom. As we will later address, this model is equivalent
to a double sine-Gordon model in the azimuthal variable $\phi$.

\subsection{Scattering phase shift method}

In this section we evaluate the products (\ref{fdet}), (\ref{pdet})
using the knowledge of bound state energies and scattering phase
shifts of the operators ${\cal H}^{s\varphi}$,  ${\cal H}^{s p}$
which have been evaluated in paper I.

In order to evaluate the density of states we consider
eigenfunctions obeying periodic boundary conditions.
The modes around the {\it metastable} state
$(\phi_m,\theta_m)=(\pi,\pi/2)$,  are then the
plane wave eigenfunction eigenfunctions of the operators
(\ref{hmop})  which can also be written as
$\sin kx$, $\cos kx$ with wavenumbers
\begin{equation}
k={2\pi n\over L}, \label{kfree}
\end{equation}
where $n=0,1,\dots$ for even parity and $n=1,2,\dots$ for
odd parity continuum eigenstates.  The corresponding density of
states is
\begin{equation}
\rho^{mi}_{(j)}= {dn\over dk}={L\over 2\pi},\qquad
i=\varphi,p\;, \quad j=e,o.
\label{rhofree}
\end{equation}

At the {\it saddle point}
$(\phi_s,\theta_s)=(\phi_s(x),\pi/2)$,  however,  we
encounter a different situation. The nonuniformity of the nucleus,
i.e. the nonconstant potentials in ${\cal H}^{s\varphi}, {\cal
H}^{s p}$  lead to  phase shifts of the continuum eigenfunctions.
In contrast to 3D problems, where the wavefunction
always vanishes at the origin, we have to distinguish
between the  phase shifts of even (e)  and odd (o) parity
wavefunctions.  We define  phase shifts as in paper I
\FL
\begin{eqnarray}
\chi_{k,(e)}^{si}(x\to\pm\infty)&\propto& \cos\left ( kx
\pm{\Delta_{(e)}^i (k)/2}\right),
\label{evenshift}\\
\chi_{k,(o)}^{si}(x\to\pm\infty)&\propto& \sin\left ( kx
\pm{\Delta_{(o)}^i (k)/2}\right),\quad
i=\varphi,p.\label{oddshift}
\end{eqnarray}
Since only the eigenfunctions of ${\cal H}^{s\;\varphi,p}$
exhibit a  phase shift, we have omitted the superscript $s$ on
$\Delta$. Note that all phase  shifts  also depend on the
parameter $R$.  Periodic boundary conditions together
with (\ref{evenshift}), (\ref{oddshift}) imply
\begin{equation}
kL+\Delta^i_{(j)} (k)=2\pi n,\quad n=1,2,\dots,
\label{period}
\end{equation}
where $i=\varphi,p$ and $j=e,o$. Following
the arguments of Ref. \onlinecite{barton},
the lowest allowed \cite{threshold} $k$-values in (\ref{period})
are $k=2\pi/L$ for odd parity eigenfunctions and $k=\pi/L$
for even parity eigenfunctions.
Note the surprising fact that the latter value does not
coincide with the lowest $k$-value (\ref{kfree}) of
even-parity solutions in the absence of a potential.

The density of states for odd-parity continuum eigenfunctions
follows from (\ref{period})
\FL
\begin{equation}
\rho^{si}_{(o)}(k)= {dn\over dk}={L\over 2\pi}+ {1\over 2\pi}
{d\Delta^i_{(o)}
(k)\over dk}, \qquad i=\varphi,p.
\label{rhoodd}
\end{equation}
Since the spectrum of even-parity continuum eigenfunctions
starts at $k=\pi/L$ while free solutions start at $k=0$,
the density of states exhibits an additional
$\delta$-function contribution at $k\to 0$
\FL
\begin{equation}
\rho^{si}_{(e)}(k)= {L\over 2\pi}+ {1\over 2\pi} {d\Delta^i_{(e)}  (k)\over
dk}-
{1\over 2}\delta(k-0^+),\quad i=\varphi,p.
\label{rhoeven}
\end{equation}
This $\delta$-function also ensures that the number
of states of the free problem equals that of the
scattering problem including bound states
\FL
\begin{equation}
\int_0^\infty\!\! dk \left [ \rho^{mi}_{(j)}-\rho^{si}_{(j)}(k)\right
]=N_{(j)}^i,\quad
i=\varphi,p,\quad j=e,o,
\label{boundcons}
\end{equation}
where, according to (I,6.25)
\begin{equation}
N^p_{(e)}=N^p_{(o)}=N^\varphi_{(o)}=1,\quad N^\varphi_{(e)}=2.
\label{nbound}
\end{equation}
are the number of even- and odd-parity bound
states of ${\cal H}^{s\varphi}$ and ${\cal H}^{sp}$.
Eq (\ref{boundcons}) with (\ref{nbound}) is verified using (\ref{rhoodd}),
(\ref{rhoeven}) and 1D Levinson's theorem (I.6.23,24) which states that
\begin{eqnarray}
&\Delta&_{(e)}^p(k=0)=\pi,\quad \Delta_{(o)}^p(k=0)=2\pi,\nonumber\\
&\Delta&_{(e)}^\varphi(k=0)=3\pi,\quad \Delta_{(o)}^\varphi(k=0)=2\pi.
\label{levtheo}
\end{eqnarray}

We are now in a position to express the ratio of the products in (\ref{fdet})
in terms of the density of states:
\FL
\begin{eqnarray}
{\prod_{k} E^{mi}_{k}
\over \prod_{k'}E^{si}_{k'} }
=\exp\left\{\int_0^{\infty}\!\!\!\!dk
\left[\rho^{mi}_{(e)}+\rho^{mi}_{(o)}-\rho^{si}_{(e)}(k)
-\rho^{si}_{(o)}(k)\right]\right.\nonumber\\
\left.\times\ln E_k^{si}\right\},
\label{ofrho}
\end{eqnarray}
where $i=\varphi,p$.
After inserting (\ref{pdet}), (\ref{rhoodd}),
(\ref{rhoeven}) into (\ref{ofrho}),
using (\ref{levtheo}) and performing a partial
integration we obtain for the $\varphi$-fluctuations
\FL
\begin{equation}
{\prod_{k} E^{m\varphi}_{k}
\over \prod_{k'}E^{s\varphi}_{k'} }
=\delta^{-6}\exp \left\{ \int_0^{\infty}\;{dk\over
2\pi}\;\left[\Delta^\varphi_{(e)}+\Delta^\varphi_{(o)}\right]
 {2k\delta^2\over 1+k^2\delta^2}\right\},
\label{of2}
\end{equation}
where we  used the fact that the phase shifts
vanish as $1/k\delta$ for $k\to\infty$ according to Born's
approximation (I,6.29).  In a completely analogous way
we obtain for the $p$-fluctuations
\FL
\begin{eqnarray}
{\prod_{k} E^{mp}_{k}
\over \prod_{k'}E^{sp}_{k'}}
&&=\left(Q^{-1}+\delta^{-2}\right)^2\nonumber\\
\times
\exp&&\left\{ \int_0^{\infty}\;{dk\over
2\pi}\;\left[\Delta^p_{(e)}+\Delta^p_{(o)}\right]
 {2k\delta^2\over 1+Q^{-1}\delta^2 +k^2\delta^2}\right\}.
\label{op2}
\end{eqnarray}
Note that in contrast to (\ref{ofrho}),
the integrands in  (\ref{of2}) and (\ref{op2})
are independent of the $k\to 0$ limit which
due to Levinson's theorem is sensitive to
the number of bound states.
This fact renders (\ref{of2}) and (\ref{op2}) suitable for
phase shift  approximations that converge nonuniformly
to the exact phase shifts  for $k\to 0$.

In the next two subsections we explicitly evaluate
the prefactor in the limit of small and large nuclei.
We show that taking the  limit $R\to 0,\infty$ of
(\ref{of2}), (\ref{op2}) is equivalent to  a direct
evaluation of the determinants of the operators that
arise in these  limits.

\subsubsection{Prefactor for $h\to 1$}

In the limit $R\to 0$, when the external field is very
close to the anisotropy field, the nucleus represents a slight
but spatially extended deviation from the metastable
state. The operators  ${\cal H}^{s\varphi}$ and
${\cal H}^{s p}$ given by (\ref{hfop}) and (\ref{hpop}) then
reduce to
\begin{eqnarray}
\bar {\cal H}^{s\varphi}&=&-{d^2\over dx^2}+\delta^{-2}
(1-6\mathop{\rm sech}\nolimits^2{x\over \delta}) \label{hfbar}\\
\bar {\cal H}^{s p}&=&-{d^2\over dx^2} +
\delta^{-2}(1-2\mathop{\rm sech}\nolimits^2{x\over \delta})+ Q^{-1}
\label{hpbar}
\end{eqnarray}
The potentials in (\ref{hfbar}), (\ref{hpbar}) are reflectionless and the
solution of the corresponding eigenvalue problems  is
well known (see also the appendix of paper I).
There are  bound states with the following energies
\begin{eqnarray}
\bar E_0^{s\varphi}&=&-3\delta^{-2},\;\;
\bar E_1^{s\varphi}=0,\nonumber \\
\bar E_0^{sp}&=&Q^{-1}.
\label{barbound}
\end{eqnarray}
Note that the two eigenvalues $E_2^{s\varphi}$, $E_1^{sp}$  of
${\cal H}^{s\varphi}$ and ${\cal H}^{s p}$  turn into
zero energy resonances\cite{zeroenergy}
of $\bar{\cal H}^{s\varphi}$ and $\bar{\cal H}^{s p}$ and
have therefore no counterparts in (\ref{barbound}).
The continuum eigenvalues of (\ref{hfbar}) and (\ref{hpbar})
are  given by  (\ref{eksp}), respectively.
Since the potentials are reflectionless the
scattering phase shifts are parity independent and given by
\begin{equation}
{\bar\Delta^\varphi}(k) =2\arctan{3 k\delta\over(k\delta)^2-2},
\label{dfbar}
\end{equation}
\begin{equation}
{\bar\Delta^p}(k)=2\arctan{1\over k\delta},
\label{dpbar}
\end{equation}
and their long wavelength behavior
\begin{equation}
\bar \Delta ^\varphi (k\to 0) = 2\pi,\;\;\; \bar \Delta ^p (k\to 0)=\pi,
\label{1dLev}
\end{equation}
is in accordance with Levinson's theorem (I,6.27) for
{\it reflectionless} potentials.
As has been discussed in paper I, Sec. VI.B,
the convergence of $\bar\Delta^\varphi$, $\bar\Delta^p$
towards the exact phase shifts $\Delta^{si}_{(j)}$
is in general nonuniform for $k=0$ (cf. Figs. I.6,7).
However, since the integrand in (\ref{of2}) and
(\ref{op2}) vanishes for $k=0$,
we can safely insert the approximations
(\ref{dfbar}) and (\ref{dpbar})
into (\ref{of2}),  (\ref{op2}), respectively,  and obtain
\begin{equation}
\lim_{R\to 0}{\prod_{k} E^{m\varphi}_{k}
\over \prod_{k'}E^{s\varphi}_{k'} }=36\,\delta^{-6},
\label{barfprod}
\end{equation}
and
\begin{equation}
\lim_{R\to 0}{\prod_{k} E^{mp}_{k}
\over \prod_{k'}E^{sp}_{k'} }=Q^{-2},
\label{barpprod}
\end{equation}
respectively.
For the evaluation of the determinants,
we substitute the bound states  (\ref{barbound}) for those in
(\ref{fdet}) and (\ref{pdet}).
However, we have
to complete (\ref{barbound}) by the  zero energy
resonances $E_2^{s\varphi}=\delta^{-2}$, $E_1^{sp}=Q^{-1}+\delta^{-2}$.
Together with (\ref{barfprod}) and (\ref{barpprod}), we then
obtain
\begin{equation}
\lim_{R\to 0}{\mathop{\rm det}\nolimits {\cal H}^{m\varphi}\over
\mathop{\rm det}\nolimits'|{\cal H}^{s\varphi}|}= 12\;\delta^{-2},
\label{fdetsmall}
\end{equation}
and
\begin{equation}
\lim_{R\to 0}{\mathop{\rm det}\nolimits {\cal H}^{mp}\over
\mathop{\rm det}\nolimits {\cal H}^{sp}}=1,
\label{pdetsmall}
\end{equation}
where in (\ref{barfprod}), (\ref{fdetsmall}), $\delta^{-2}= R^2$.
The  result (\ref{pdetsmall}) is remarkable since it shows
that the fluctuations in $p$-direction do not play a
role at all for small nuclei,  independent
of the size of the hard-axis anisotropy constant.
This suggests that in the limit $R\to 0$ the system may effectively be
described by a  double sine-Gordon equation in the azimuthal angle $\phi$.
We shall return to this issue in Sec. VII.

Alternatively, although less carefully, we can  interchange the
limit $R\to 0$ with the functional integration,  and directly calculate
\begin{eqnarray}
{\mathop{\rm det}\nolimits {\cal H}^{m\varphi}\over
\mathop{\rm det}\nolimits' |\bar{\cal H}^{s\varphi}|}
&=&{\prod_{k}  E^{m\varphi}_{k}
\over  |\bar E_0^{s\varphi}|  \prod_{k'}\bar E^{s\varphi}_{k'}},
\label{fdetbar}\nonumber\\
{\mathop{\rm det}\nolimits {\cal H}^{m p}\over
\mathop{\rm det}\nolimits \bar {\cal  H}^{sp}}
&=&{\prod_{k} E^{mp}_{k}
\over \bar E_0^{sp}  \prod_{k'} \bar E^{sp}_{k'}}.
\label{pdetbar}
\end{eqnarray}
For the evaluation of the r.h.s. in (\ref{pdetbar}) we  proceed
similarly to the derivation of (\ref{of2}) and (\ref{op2}) with the
following modifications:
The density of states $\rho^{mi}_{(j)}-\rho^{si}_{(j)}(k)$ in
(\ref{ofrho}) has to be replaced by
$\bar\rho^{mi}_{(j)}-\bar\rho^{si}_{(j)}(k)=-{1\over 2\pi} d\bar \Delta^i
/dk$
with $\bar \Delta^i$ given by (\ref{dfbar}) and
(\ref{dpbar}) ($i=\varphi,p$). Using the version
(\ref{1dLev}) of Levinson's theorem together with
(\ref{dfbar}) and (\ref{dpbar}), we recover the results
(\ref{fdetsmall}) and (\ref{pdetsmall})  after integration.

To summarize, we thus have shown that the
small nucleus approximation
may be used for the evaluation of the fluctuation
determinants, or explicitly
\begin{eqnarray}
\lim_{R\to 0} {\mathop{\rm det}\nolimits {\cal H}^{m\varphi}\over
\mathop{\rm det}\nolimits' |{\cal H}^{s\varphi}|} = {\mathop{\rm det}
\nolimits {\cal H}^{m\varphi}\over
\mathop{\rm det}\nolimits' |\bar{\cal H}^{s\varphi}|},\nonumber\\
\lim_{R\to 0} {\mathop{\rm det}\nolimits {\cal H}^{mp}\over
\mathop{\rm det}\nolimits {\cal H}^{sp}} = {\mathop{\rm det}\nolimits
{\cal H}^{mp}\over
\mathop{\rm det}\nolimits \bar{\cal H}^{sp}}.
\label{detlim}
\end{eqnarray}
where the r.h.s. has been evaluated in leading order in $R$.
The relation (\ref{detlim}) has not been clear on the onset,
since the operators $\bar{\cal H}^{s\varphi}$,
$\bar{\cal H}^{sp}$ exhibit a different
number of bound states and different long wavelength behavior of
the scattering phase shifts than the operators ${\cal H}^{s\varphi}$,
${\cal H}^{sp}$. As we have shown now, these two differences
conspire in such a way that an interchange of the limits
in (\ref{detlim}) is indeed correct.

\subsubsection{Prefactor for $h\to 0$}

For $R\gg 1$ the nucleus separates into two independent $\pi$-Bloch walls.
Correspondingly  ${\cal H}^{s\varphi}$ and ${\cal H}^{s p}$ merge into
the same operator $\hat
{\cal H}^s$ which consists of two independent potential wells of the form
$-2\delta^{-2}\mathop{\rm sech}\nolimits^2(x/\delta\pm R)$.
The bound states of $\hat{\cal H}^s$  are then
given by the symmetric and antisymmetric linear combinations of
the ground states of the single wells and have energies (cf. (I,6.4,6.6))
\begin{eqnarray}
\hat E_0^{sp}&=&Q^{-1},\quad \hat E_1^{sp}=Q^{-1}+8 e^{-2R},\nonumber\\
\hat E_0^{s\varphi}&=&-8e^{-2R},\quad \hat E_1^{s\varphi}=0,
\label{hatbound}
\end{eqnarray}
Note, that in this approximation $E_2^{s\varphi}$ has merged into
a  zero energy resonance of $\hat {\cal H}^s$. The continuum
eigenvalues are identical to (\ref{eksf}) while the
scattering phase shifts are twice those of
a single potential well (cf. (\ref{dpbar}))
\begin{equation}
\hat\Delta^s (k)=4 \arctan{1\over k\delta}.
\label{hatd}
\end{equation}
The coincidence of the phase shifts of even and odd eigenfunctions
originates from the fact that the two $-2 \mathop{\rm sech}\nolimits^2 x$
potential wells are reflectionless. The phase shifts (\ref{hatd})
obey the reflectionless version of Levinson's theorem, i.e.
$\hat Delta^s(k\to 0) =\pi N$ with $N$ the number of bound
states. But as in the previous
subsection,  the phase shifts $\Delta^{s\varphi}_{(e)}$ and
$\Delta^{sp}_{(e)}$ only converge on the open interval $0<k<\infty$
towards $\hat\Delta^s$. Inserting (\ref{hatd}) into
(\ref{of2}),(\ref{op2}), and using (\ref{hatbound}) together with
$E_2^{s\varphi}=\delta^{-2}$ in (\ref{fdet}), (\ref{pdet}), we
obtain
\begin{equation}
\lim_{R\to\infty}{\mathop{\rm det}\nolimits {\cal H}^{m\varphi}
\over \mathop{\rm det}\nolimits'|{\cal H}^{s\varphi}|}=2 e^{2R},
\label{fdetlarge}
\end{equation}
\begin{equation}
\lim_{R\to\infty}{\mathop{\rm det}\nolimits {\cal H}^{mp}\over
\mathop{\rm det}\nolimits {\cal H}^{sp}}=[\sqrt{1+Q} +\sqrt{Q}]^4.
\label{pdetlarge}
\end{equation}
The latter result approaches $1$ for large
hard-axis anisotropies as expected.  However, in the opposite limit of
high $Q$, the $p$-fluctuations  lead to a prefactor (\ref{pref})
proportional to $Q$.

\subsubsection{Out of easy plane fluctuations for $Q^{-1}\delta^2\to\infty$}

A large hard-axis anisotropy leads to the suppression of out of easy-plane
fluctuations by the existence of a large mass gap. Therefore
we expect the fluctuation determinant to become one in this limit.

To prove this conjecture, we remark that
$\Delta_{(e)}^{sp}$, $\Delta_{(o)}^{sp}$ are continuous
functions which are proportional to $1/k\delta$ for $k\to\infty$
according to Born's approximation (I,6.29) and remain
finite for $k\to 0$ due to Levinson's theorem. Therefore
both phase shifts obey the inequality $\Delta(k)<c/k\delta$ with
a suitably chosen constant $c$.  For the integral in (\ref{op2})
we thus obtain the following inequality
\FL
\begin{equation}
0<\int_0^{\infty}\!{dk\over
\pi}\;{k\delta^2\left[\Delta^p_{(e)}+\Delta^p_{(o)}\right]
 \over 1+Q^{-1}\delta^2 +k^2\delta^2}
<{ d \over \sqrt{1+Q^{-1} \delta^2}},
\label{largeQest}
\end{equation}
where $d=(c_{(e)}^p+c_{(o)}^p)/2$.
Since $d$ is independent of $Q$, the upper limit tends to
zero for $Q^{-1}\delta^2\to\infty$ and therefore
we have with $E_0^{sp}\equiv Q^{-1}$, $E_1^{sp}=Q^{-1}+\varepsilon
\delta^{-2}$ with $0<\varepsilon<1$ and (\ref{op2}), (\ref{pdet})
\begin{equation}
\lim_{Q^{-1}\delta^2\to\infty}{\mathop{\rm det}
\nolimits{\cal H}^{mp}\over \mathop{\rm det}\nolimits {\cal H}^{sp}}=1.
\label{pdet1}
\end{equation}
In fact we have now shed some new light on the result (\ref{pdetsmall}).
The hard-axis anisotropy does not enter in isolated form but rather
in the combination $Q^{-1}\delta^2=Q^{-1}\coth^2 R$ which shows that
due to the diverging length scale, the hard-axis anisotropy becomes
effectively strong for small $R$, no matter how small $Q^{-1}$ is.

\subsection{Jacobi method}

There is also an alternative way for evaluating the products of
eigenvalues which has its origin in the space slice representation
of the path integrals.
This method allows for the exact evaluation of the statistical
prefactor in the $\varphi$-variable for all values of $R$. In the limit
$Q^{-1}=0$ we are also able to evaluate the prefactor for
out of easy-plane fluctuations.
We first show how this method can be applied
to the evaluation of (\ref{fdet}). According to Ref.\onlinecite{Schulman}
we have
\begin{equation}
{\mathop{\rm det}\nolimits_L{\cal H}^{m\varphi}
\over \mathop{\rm det}\nolimits_L{\cal H}^
{s\varphi}}={D_\varphi^{(0)}(L/2) \over
D_\varphi(L/2)} \label{detL} \end{equation}
The notation $\mathop{\rm det}\nolimits_L$ on the l.h.s. of
(\ref{detL}) indicates that the evaluation of the determinants relies on
eigenvalue problems defined on the finite interval
$[-L/2 , L/2]$ with respect to functions that vanish at the end of the
interval.
The functions $D_\varphi$ and $D_\varphi^{(0)}$ on
the r.h.s. of (\ref{detL})  are  solutions of the
differential equations
\begin{eqnarray}
{\cal H}^{s\varphi} D_\varphi(x) =0,\label
{Df}\\
\left(-{d^2\over dx^2}+\delta^{-2}\right)D_\varphi^{(0)}(x)=0,\label{Df0}
\end{eqnarray}
with the following ``initial" conditions (the prime denotes $d/dx$)
\begin{eqnarray}
D_\varphi (-L/2)&=&0,\quad  D_\varphi'(-L/2)=1\label{Dfbc}\\
D_\varphi^{(0)}(-L/2)&=&0,\quad  {D^{(0)}_\varphi}'(-L/2)=1.
\label{Df0bc}
\end{eqnarray}

Note that on the finite interval the first excited eigenfunction of
${\cal H}^{s\varphi}$ has no longer zero energy
and therefore the l.h.s of (\ref{detL}) is well defined.
The eigenvalue problem of this quasi zero-energy mode is written as
\begin{equation}
{\cal H}^{s\varphi} f= \mu f,
\label{fev}
\end{equation}
where for large system lengths $L$, $\mu>0$ is small, and
$f$ obeys the boundary conditions $f(\pm L/2)=$$f'(\pm L/2)=0$.
Note that for $L\to\infty$
we have $f\to\chi^{s\varphi}_1$ and $\mu\to 0$. The fluctuation
determinant  is then obtained as follows:
\begin{equation}
{\mathop{\rm det}\nolimits {\cal H}^{m\varphi}\over\mathop{\rm det}
\nolimits'|{\cal H}^{s\varphi}|}= \lim_{L\to\infty}\left| \mu\
 {D_\varphi^{(0)}(L/2) \over D_\varphi(L/2)}\right|.
\label{detL2}
\end{equation}
We now turn to the evaluation of the r.h.s. of (\ref{detL2}).
$D_\varphi^{(0)}(L/2)$ is easily obtained  by integration of (\ref{Df0}) with
(\ref{Df0bc})
\begin{equation}
D_\varphi^{(0)}(L/2)=\delta \sinh(L/\delta)\simeq{\delta\over 2}e^{L/\delta}.
\label{Df0asy}
\end{equation}

The function $D_\varphi(x)$ obeys the same
differential equation (\ref{Df}) as the zero mode $\chi_1^{s\varphi}(x)$
(\ref{cf1}), but
subject to different boundary conditions (\ref{Dfbc}).
Therefore $D_\varphi(x)$ is a linear
combination of $\chi_1^{s\varphi}(x)$ and the unknown linear independent
solution $\xi_1^{s\varphi}(x)$ of the differential equation (\ref{Df})
\begin{equation}
D_\varphi(x)=u\chi_1^{s\varphi}(x)+v\xi_1^{s\varphi}(x),\label{Dflin}
\end{equation}
with real constants $u, v$. $\chi_1^{s\varphi}$ is taken to be unnormalized
\begin{equation}
\chi_1^{s\varphi}={d\phi_s\over dx}=
\delta^{-1}\left[\mathop{\rm sech}\nolimits(x/\delta-R)-\mathop{\rm
sech}\nolimits(x/\delta+R)\right]. \label{zeromode}
\end{equation}
The normalization of $\xi_1^{s\varphi}(x)$
is chosen such that the Wronskian
\begin{equation}
\chi_1^{s\varphi}{\partial \xi_1^{s\varphi}\over\partial x}
-\xi_1^{s\varphi}{\partial \chi_1^{s\varphi}\over\partial x}=1.
\label{wronski}
\end{equation}
In order to satisfy the initial conditions (\ref{Dfbc}), we must have
\begin{equation}
u=-\xi_1^{s\varphi}(-L/2),\qquad v= \chi_1^{s\varphi}(-L/2).
\label{uv}
\end{equation}
{}From (\ref{zeromode}) we infer the asymptotic behaviour
\begin{equation}
\chi_1^{s\varphi}\to N {\rm sgn} (x) e^{-|x|/\delta},
\quad \hbox {for } x\to\pm \infty,
\label{chiasy}
\end{equation}
with
\begin{equation}
N=4\delta^{-1}\sinh R.
\label{N}
\end{equation}
The symmetry of the potential
in ${\cal H}^{s\varphi}$ and the antisymmetry
of $\chi_1^{s\varphi}$ allow us to choose  $\xi_1^{s\varphi}$
as a symmetric function which  has the
asymptotic behaviour
\begin{equation}
\xi_1^{s\varphi}\to N'e^{|x|/\delta},
\quad \hbox{for } x\to\pm \infty.
\label{xiasy}
\end{equation}
{}From (\ref{wronski}) it follows that
\begin{equation}
N'={\delta\over 2 N},
\label{Nprime}
\end{equation}
and therefore with (\ref{uv})-(\ref{Nprime}) we have
\begin{equation}
D_\varphi(L/2)=-\delta.
\label{Dfasy}
\end{equation}

Finally we are left with the evaluation of $\mu$.
The eigenfunction
$f$  may to first order in
$\mu$ be expressed as
\begin{equation}
f(x)=\eta(x)+\mu\int_{-L/2}^{L/2}G(x,y)\eta(y) dy,
\label{f}
\end{equation}
with the Green's function
\begin{equation}
G(x,y)=\theta(x-y)[
\chi_1^{s\varphi}(x)\xi_1^{s\varphi}(y)
-\chi_1^{s\varphi}(y)\xi_1^{s\varphi}(x)].
\label{G}
\end{equation}
The quasi zero-energy eigenvalue $\mu$ is now
determined such that $f(\pm
L/2)=0$.  The function $\eta$ is a solution
of the homogeneous problem (\ref{Df}) which  satisfies
the boundary condition $\eta(-L/2)=0$,
\begin{equation}
\eta(x)=\chi_1^{s\varphi}(x)+c\xi_1^{s\varphi}(x),
\label{eta}
\end{equation}
with
\begin{equation}
c=-\chi_1^{s\varphi}(-L/2)/\xi_1^{s\varphi}(-L/2).
\label{c}
\end{equation}
The requirement $f(L/2)=0$ then  leads to
\FL
\begin{equation}
\mu={\chi_1^{s\varphi}(L/2)+c\xi_1^{s\varphi}(L/2)
\over \int_{-L/2}^{L/2}dy[(\chi_1^{s\varphi}(y))^2\xi_1^{s\varphi}(L/2)
-c\chi_1^{s\varphi}(L/2)(\xi_1^{s\varphi}(y))^2]}.
\label{mu}
\end{equation}
Since the normalization of $\chi^{s\varphi}_1$ is
independent of $L$, the first term in the denominator is of the order
$\exp(L/2\delta)$ whereas the second vanishes as $\exp(-L/2\delta)$
and thus may be neglected. In leading order in $\exp(-L/\delta)$
we thus obtain
\begin{equation}
\mu={2\chi_1(L/2)\over
\xi_1(L/2) \int_{-L/2}^{L/2} \chi_1^2(y)}={64\;\delta^{-3}\sinh^2 R
\over  {\cal E}_s(R)} e^{-L/\delta},
\label{mures}
\end{equation}
where we have made use of (\ref{zeromode}) and (I,3.12).
Inserting (\ref{Df0asy}), (\ref{Dfasy}) and (\ref{mures})
into (\ref{detL2}) and performing the limit $L\to\infty$,
we finally obtain the result
\begin{equation}
{\mathop{\rm det}\nolimits {\cal H}^{m\varphi}\over\mathop{\rm det}
\nolimits'|{\cal H}^{s\varphi}|}={8\tanh^3 R\sinh^2 R\over
                  \tanh R-R \mathop{\rm sech}\nolimits^2 R}.
\label{detfex}
\end{equation}
which is {\it exact} for all values of the external field.
Note that in the limits $R\to0$ and $R\to  \infty$, Eq. (\ref{detfex}) reduces
to the results (\ref{fdetsmall}), (\ref{fdetlarge}), respectively.

The above method cannot be used for the evaluation of the $p$-determinants
(\ref{omfexact}) for arbitrary values of the hard-axis anisotropy since the
zero energy eigenfunction of ${\cal H}^{sp}$ is not explicitly known.
In the limit of small $Q^{-1} \delta^2$, however, the fluctuation
determinant may be calculated exactly:

The $Q^{-1}$-independent operator
${\cal H}^{sp}-Q^{-1}$  exhibits the
zero energy mode $\chi_0^{sp}$ and we can proceed along the same lines
\cite{detcomment} as in the derivation of (\ref{detfex}) to obtain
\begin{equation}
{ \mathop{\rm det}\nolimits ( {\cal H}^{mp}-Q^{-1} ) \over
\mathop{\rm det'}\nolimits ( {\cal H}^{sp}-Q^{-1} )} =
{8\tanh^3 R\cosh^2 R\over
                     \tanh R+R \mathop{\rm sech}\nolimits^2 R},
\label{detpex}
\end{equation}
where the prime denotes omission of the zero energy mode.
{}From (\ref{ofrho})  with (\ref{eksp})   it follows that
$\prod_k E_k^{mp}/\prod_k' E_k'^{sp}=$ $\prod_k (E_k^{mp}-Q^{-1})/
\prod_k'  (E_k'^{sp}-Q^{-1})+{\cal O}(Q^{-1}\delta^2)$.
With (\ref{pdet}) we then obtain in leading order in the
small parameter $Q^{-1} \delta^2$
\begin{eqnarray}
{ \mathop{\rm det}\nolimits  {\cal H}^{mp} \over
\mathop{\rm det}\nolimits  {\cal H}^{sp}}
&=& {1\over E_0^{sp} } {E_1^{sp} -Q^{-1} \over E_1^{sp} }
{\mathop{\rm det}\nolimits  ({\cal H}^{mp}-Q^{-1} ) \over
\mathop{\rm det'}\nolimits ( {\cal H}^{sp}-Q^{-1} )} \nonumber \\
&=&Q {E_1^{sp} -Q^{-1} \over E_1^{sp} }
{8\tanh^3 R\cosh^2 R\over
\tanh R+R \mathop{\rm sech}\nolimits^2 R}.
\label{pdetQlarge}
\end{eqnarray}
The second factor on the r.h.s has been retained since
the coefficients of its power expansion in $Q^{-1}\delta^2$
diverge for $R\to \infty$. However, in the limit
$Q^{-1}\ll E_1^{sp}$ ($R$ not too large),
it reduces to one. Note that the result (\ref{pdetQlarge})
is only valid for $Q^{-1} \delta^2$ small and
it can therefore not be used for $R$ small. In this latter
case (\ref{pdet1}) applies.

With the exact result (\ref{detfex}), the
prefactor (\ref{pref}) now takes the following form
\begin{equation}
\Omega=\lambda_+ L \sqrt{\beta {\cal A}} {4\over \pi^{3/2}}
\tanh^{3/2}R\; \sinh R
\sqrt{  {   \mathop{\rm det}\nolimits {\cal H}^{mp}\over
\mathop{\rm det}\nolimits {\cal H}^{sp}   }    },
\label{omfexact}
\end{equation}
which is further evaluated in the following section.

\section{Nucleation Rates in the overdamped limit}

 In this section we shall derive analytical results for the prefactor
(\ref{omfexact}) in the limit of large and small values of $Q^{-1}\delta^2$
as well  as large $R$.  In the intermediate parameter range
the prefactor is evaluated numerically.
The discussion in this section is
restricted to the regime of large damping . The case of  moderate
damping which is more relevant for  real systems shall
be discussed  in the next section.

We start with the evaluation of the
decay frequency $\lambda_+$ of the nucleus.
For large values of the damping constant $\alpha$, the equations
(\ref{lambda}) characterizing the unstable mode of the nucleus decouple
and take the form
\begin{eqnarray}
\lambda_+\varphi_+&=-\alpha{\cal H}^{s\varphi}\varphi_+,
\nonumber\\
\lambda_+ p_+&=-\alpha{\cal H}^{sp} p_+,
\label{instablarge}
\end{eqnarray}
with $\lambda_+>0$.  The unstable mode is thus given by
the ground state of ${\cal H}^{s\varphi}$,
i.e. $(\varphi_+,p_+)\propto (\chi^{s\varphi}_0,0)$ and for large
$\alpha$ the corresponding escape frequency is given by
\begin{equation}
\lambda_+=\alpha |E_0^{s\varphi}(R)|.
\label{lambdalarge}
\end{equation}
The eigenvalue  $E_0^{s\varphi}(R)$ has been investigated  in paper I.
It is known analytically in the limits $R\to 0,\infty$
(cf. (\ref{barbound}), (\ref{hatbound}) respectively)
which allows an explicit evaluation of the prefactor in
these limits.

i) For large nuclei ($R\to\infty$) we may use  (\ref{hatbound}),
(\ref{pdetlarge}),  and together with
${\cal E}_s\to 4$ and the  prefactor (\ref{omfexact})
takes the following form
\begin{equation}
\Omega=\alpha L\sqrt{\beta{\cal A}}{16\over\pi^{3/2}}
\left[\sqrt{Q}+\sqrt{1+Q}\right]^2 e^{-R}.
\label{omlargeR}
\end{equation}

ii) For $Q^{-1}\delta^2\gg 1$, (\ref{pdet1}) and
(\ref{lambdalarge}) may be inserted into (\ref{omfexact}) to yield
\begin{equation}
\Omega=\alpha L\sqrt{\beta{\cal A}}{4\over\pi^{3/2}}
|E_0^{s\varphi}(R)| \tanh^{3/2}\!R\; \sinh R.
\label{omQsmall}
\end{equation}
For small $R$, this reduces to
\begin{equation}
\Omega=\alpha L\sqrt{\beta{\cal A}}{12\over\pi^{3/2}}
R^{9/2}.
\label{omsmallR}
\end{equation}
Note that the results, (\ref{omQsmall}) and (\ref{omsmallR})
are both independent of the value of the hard-axis anisotropy.

iii) In the limit $Q^{-1} \delta^2 \to 0$  we can use
(\ref{omfexact}) and (\ref{pdetQlarge})  to obtain
\FL
\begin{eqnarray}
\Omega=\alpha L\sqrt{\beta{\cal A}}{8\sqrt{2}\over
\pi^{3/2}} |E_0^{s\varphi}|  \sqrt{Q}
\sqrt{{  E_1^{sp}-Q^{-1}\over E_1^{sp}  }} \nonumber\\
\times{\tanh^2 R \sinh^2 R \over
\sqrt{\tanh R+ R\mathop{\rm sech}\nolimits^2 R}}.
\label{Qinfty}
\end{eqnarray}
For values of $R$ such that $Q^{-1}\ll E_1^{sp}(R)$, the
square root containing $E_1^{sp}$ reduces to 1.
In the limit $R\to\infty$, Eqn (\ref{Qinfty}) reduces to
(\ref{omlargeR}) with $Q^{-1}\to 0$

Note however, that the present theory is only valid
for hard-axis anisotropies which are not too small
such that the amplitude of out of easy-plane fluctuations
is much smaller than one. Requiring the thermal expectation
value $\langle p^2 \rangle$ with respect to
the Boltzmann weight $\exp\{-\beta {\cal A}{\cal E}^{(2)}_s\}$
with ${\cal E}^{(2)}_s$ as in (\ref{e2nucl})
to be smaller than one and noting that the lowest
eigenvalue in $p$-direction is $E_0^{sp}=Q^{-1}$,
we obtain for the validity of the
present theory the condition
$\beta {\cal A} \sqrt{A K_e} K_h/K_e >1$.

For all other parameter values the prefactor has been
evaluated numerically. The corresponding results are shown in Fig. \ref{fig3}
with the dashed and dashed-dotted lines representing the
asymptotic formulas (\ref{omlargeR})  and
(\ref{omsmallR}) respectively.
Reinstating the units (\ref{units}) we recognize that
the prefactor (\ref{pref}) is inversely proportional
to $\alpha$, a fact which is  in accordance with the
general behavior \cite{hanggi} of nucleation rates in
the overdamped limit.

\section{Nucleation Rate for Moderate Damping}

In the last section we have presented results for nucleation
rates of thermally activated magnetization reversal in the overdamped limit.
In that case, the decay of  the nucleus is governed by purely
dissipative mechanisms. However,  in ferromagnetic
materials the damping constant is usually of the order $\alpha=10^{-2}$
or sometimes as small as $10^{-4}$ in some high purity materials such
as YIG.

According to (\ref{pref}), the dynamic properties of the
system enter the nucleation rate only \cite{underdamped} in the form
of the decay frequency of the nucleus.
In order to evaluate the nucleation rate in
the moderately damped regime  we therefore have to
include the conservative,  precessional  part of the dynamics
for the evaluation of the decay frequency $\lambda_+$.
In the limits of small and large nuclei this  decay frequency
can again be expressed in closed analytical form thus enabling us to
give exact results of the total prefactor and hence of the
total rate.
To the best of my knowledge, this provides the first application
of Langer's general theory \cite{Langer} of moderate friction
to a system with infinitely many degree of freedoms.

In a first part we discuss the escape frequency $\lambda_+$.
We obtain exact expressions in the limits $R\to \infty$ and
$Q^{-1}\delta^2\to \infty$ as well as an approximate
formula which expresses
$\lambda_+$ by $E_0^{s\varphi}$.
These results allow an exact evaluation of the  nucleation rate
in the limits $R\to 0$  and $R\to\infty$.

\subsection{Escape Frequency}

The dynamically unstable mode $(\varphi_+,p_+)$ of the nucleus is
the solution of the (non-Hermitian) coupled eigenvalue problem (\ref{lambda}).
Before turning to a  quantitative analysis
we give a qualitative discussion of the parity and relative sign
of the functions $\varphi_+$, $p_+$.
Both functions are nodeless and symmetric in $x$ with opposite
relative sign as may be seen from the following plausibility arguments:
For $\alpha\to\infty$ we know from (\ref{instablarge})
that the dynamical unstable mode
coincides with the ground state of ${\cal H}^{s\varphi}$, i.e.
$(\varphi_+,p_+)\propto (\chi^{s\varphi} _0,0)$, and therefore
$\varphi_+$ is a symmetric nodeless function
in $x$. For finite values of $\alpha$ there will be a nonzero $p_+$-component.
Since the nucleus represents an untwisted
$\pi$-Bloch wall pair (cf. (I, 3.11 )), the instability represents
a confluence or a  separation of the two domain walls which
is associated with a monotonical increase or decrease of the
angle $\phi$. Hence $\varphi_+$ is a symmetric nodeless function in $x$.
To comment on $p_+$, we have to recall that a motion of the domain
wall is only possible if the structure exhibits an out of easy-plane
component \cite{braunbrod}. In order for the domain walls to move in opposite
directions, the out of easy-plane component must have the same sign
at the center of the two oppositely twisted kinks
and due to the gyroscopic nature of the equations of motion, $p_+$
and $\varphi_+$ must have opposite signs.

Therefore we are looking for even-parity nodeless
solutions of (\ref{lambda}) with opposite signs.
The ambiguity in the overall sign of $(\varphi_+,p_+)$ describes
the freedom of the nucleus either to collapse or to expand.
Inspecting (\ref{lambda})  we recognize that the eigenvalue problem
can be easily solved if $\varphi_+$ and $p_+$ are  the ground states
of ${\cal H}^{s\varphi}$ and  ${\cal H}^{sp}$ respectively and
proportional to each other. This is fulfilled in two limiting cases,
i) $R$ large and ii) $Q^{-1}\delta^2\gg 1$.

i) For large $R$ we have according to (\ref{cp1}) and
   (I, 6.2)
\begin{equation}
\chi^{s\varphi} _0\propto\chi^{sp} _0\propto\mathop{\rm sech}
\nolimits (x/\delta +R) + \mathop{\rm sech}\nolimits(x/\delta -R).
\label{largeRef}
\end{equation}
Inserting $\varphi_+\propto$ $ p_+\propto \chi_0^{s\varphi}$
into (\ref{lambda}) and using that the ground state energy
of ${\cal H}^{sp}$ is given by $E_0^{sp}\equiv Q^{-1}$ we obtain
\FL
\begin{eqnarray}
 \lambda_+ =&-&{\alpha\over 2}[Q^{-1} + E^{s\varphi}_0]\nonumber\\
&+&\sqrt{ \left({\alpha\over2}
\right)^2[Q^{-1}-E^{s\varphi}_0]^2 - Q^{-1}E^{s\varphi}
_0},
\label{lalargeR}
\end{eqnarray}
where $E^{s\varphi} _0\simeq -8 e^{-2R}$ in the limit
$R\to \infty$.
The square root in (\ref{lalargeR}) has been retained because the relative
magnitude of the (small) parameters $E_0^{s\varphi}$, $\alpha$,
and  $Q^{-1}$ has not been specified yet.
The corresponding unstable mode is given by
\begin{eqnarray}
(\varphi_+,p_+)\propto
\left[\mathop{\rm sech}\nolimits ({x\over\delta }+R) +
\mathop{\rm sech}\nolimits({x\over\delta}
-R)\right]\nonumber\\
\times\left( 1, - Q[\lambda_++\alpha E^{s\varphi} _0]\right).
\label{modelargeR}
\end{eqnarray}
The plus sign of the square
rooot is chosen in order to reproduce the correct asymptotic
behavior (\ref{lambdalarge}) for large $\alpha$.
Note that Eq. (\ref{modelargeR}) agrees with the
statements made above: The functions $p_+, \varphi_+$ are
symmetric and nodeless while the ratio
$p_+/\varphi_+$ is always negative and
vanishes for $\alpha\to\infty$.
For $\alpha\to 0$ we have
$p_+/\varphi_+=-\sqrt{Q |E^{s\varphi} _0(R)|}$.

ii) For $Q^{-1}\delta^{2}\gg 1$ we have ${\cal H}^{s p}p_+ =Q^{-1}p_+ +{\cal O
}(\delta^{-2})$. Thus  the first Eq. of  (\ref{lambda}) can be solved
 for $p_+$ and after insertion into the second eq. of
(\ref{lambda}), the following
eigenvalue problem is obtained:
\begin{equation}
{\cal H}^{s\varphi} \varphi_+=-\lambda_+{\lambda_+ Q+\alpha\over 1+
\alpha^2+\alpha\lambda_+ Q}\;\varphi_+,
\label{Hev}
\end{equation}
The solution of this equation is known, i.e. $\varphi_+\propto
\chi^{s\varphi} _0$,
and hence the coefficient of $\varphi_+$ on the r.h.s.  equals
$E^{s\varphi}_0$.
Solving for $\lambda_+$ we recover the
expressions (\ref{lalargeR}) for $\lambda_+$ and
(\ref{modelargeR}) for $p_+/\varphi_+$
but with $E_0^{s\varphi}(R)$ now evaluated
for arbitrary values of $R$.
This is a remarkable result  as it demonstrates the validity of
(\ref{lalargeR}), (\ref{modelargeR}) in the opposite limits
$Q^{-1}\delta^2\gg 1$, $R\gg 1$. Note that for large
$Q^{-1}$, (\ref{lalargeR}) and (\ref{modelargeR}) hold for all values
of $R$. In the particular case of small $R$ ($\delta^2 \gg 1$)
we can insert the small $R$ approximation
$E^{s\varphi} = -3 R^2+ {\cal O} (R^4)$ into  (\ref{lalargeR}) to obtain
\begin{eqnarray}
 \lambda_+= &-&{\alpha\over2}[Q^{-1}-3 R^2]+\nonumber\\
&+&\sqrt{\left({\alpha\over2}\right)^2
[Q^{-2}+6 Q^{-1} R^2]+ 3 Q^{-1}
R^2}.
\label{lasmallR}
\end{eqnarray}
The unstable mode is then given by
\begin{equation}
(\varphi_+,p_+) \propto \mathop{\rm sech}\nolimits ^2 ({x\over\delta} )
( 1, - Q(\lambda_+-3\alpha R^2)).
\end{equation}
The validity of the expression (\ref{lalargeR}) in the opposite
limits $R\to 0,\infty$
might hint to a more extended validity.
In order to investigate $\lambda_+$ for intermediate values
of $R$ at arbitrary $Q^{-1}$ we have to resort to numerical methods.
It turns out that the
direct integration of (\ref{lambda}) yields rather inaccurate
results (errors of 10\% ). Considerable improvement has been achieved
by converting (\ref{lambda}) into two decoupled
$4^{\rm th}$ order differential equations in each of the variables
$\varphi_+,p_+$.
As is seen from Fig. \ref{fig4},
Eq. (\ref{lalargeR}) provides an excellent
approximation to these numerical results.

\subsection{Nucleation Rates}

We are now in a position to give the results for the prefactor for
moderate damping. Results in closed form are obtained in the
limits $R\to 0$ and $R\to \infty$.  For $Q^{-1}\delta^2\to \infty$
the prefactor can be  expressed in terms of the negative
eigenvalue $E_0^{s\varphi}$ which
for arbitrary values of $R$ has to be evaluated numerically.

i) For large $R$, we can combine   (\ref{omfexact}),
(\ref{pdetlarge}) and (\ref{lalargeR}) with
$E_0^{s\varphi}=-8 e^{-2R}$ to obtain
\begin{eqnarray}
\Omega&=&L\sqrt{\beta {\cal A}} {2\over \pi^{3/2}}
[\sqrt{Q}+\sqrt{1+Q}]^2 e^R \left\{ -{\alpha\over 2} (Q^{-1}-8e^{-2R})\right.
\nonumber\\
&&\left.+ \sqrt{\left( {\alpha\over 2}^2\right)
  [Q^{-2}+16Q^{-1}e^{-2R}] +8Q^{-1}e^{-2R} } \right\}.
\label{preflargeR}
\end{eqnarray}
The square root has been retained since the relative order of the
small parameters $\alpha$ and $e^{-2R}$ has not been specified yet.
However, in an expansion of the square root
only leading terms in $e^{-2R}$ should be
taken into account.

ii) For $Q^{-1}\delta^2\to \infty$, the results (\ref{omfexact}) and
(\ref{pdet1}) yield
\begin{equation}
\Omega= \lambda_+ L\sqrt{\beta{\cal A}}{4\over \pi ^{3/2}}
\tanh^{3/2}R\; \sinh R
\label{smallRex}
\end{equation}
with $\lambda_+$  given by (\ref{lalargeR}).  For small $R$, this
result reduces with (\ref{lasmallR}) to
\begin{eqnarray}
\Omega&=&L\sqrt{\beta {\cal A}} {4\over \pi^{3/2}} R^{5/2}
\left\{ -\alpha(Q^{-1}-3R^2)
\phantom{\sqrt{ \left( {\alpha\over 2}\right)^2     }   }\right.
\nonumber\\
&&\left.+ \sqrt{\left( {\alpha\over 2}\right)^2
  [Q^{-2}+6Q^{-1}R^2] +3Q^{-1}R^2 } \right\}.
\label{prefsmallR}
\end{eqnarray}
Also here, the square root has been retained since we did not specify the
relative magnitude of the small parameters $\alpha$ and
$R$. However, we have to keep in mind that upon expansion of the
square root only terms in leading order in $R$ have to be kept
in order to be consistent with the derivation of the
determinant of $p$-fluctuations.
For small damping constants $\alpha\ll\sqrt{Q} R$, Eq. (\ref{prefsmallR})
reduces to
\begin{equation}
\Omega=L\sqrt{\beta{\cal A}}{4\sqrt{3}\over \pi^{3/2}}
\sqrt{Q^{-1}} R^{7/2}.
\label{prefalR}
\end{equation}
This limit is realized in typical experimental  situations.
For $\sqrt{Q}R\ll \alpha$ we obtain
\begin{equation}
\Omega=L\sqrt{\beta {\cal A}}{12\over \pi^{3/2}} \left( \alpha+{1\over
\alpha}\right) R^{9/2}.
\label{prefRla}
\end{equation}
For large values of $\alpha$, Eq. (\ref{prefRla}) merges into the
prefactor for the overdamped regime (\ref{omsmallR}).

The  above results for the prefactor are summarized in
table I. In Fig. 5, numerical results for the prefactor are shown
for arbitrary values of $R$. The prefactor
is maximal for $R\simeq 1$ and decreases as
the external field approaches the
anisotropy field, i.e $h\to 1$, or as the field
approaches zero. One clearly recognizes that
$\Omega$ is independent of the hard-axis anisotropy $Q^{-1}$ and
the damping constant $\alpha$ for small and intermediate
values of $R$, respectively  as predicted by  (\ref{prefRla}), (\ref{prefalR}).

The total rate for magnetization reversal
is then given by (\ref{rate}) and  experimental
consequences of this result shall be discussed in section VIII.
A detailed discussion of experimental implications of these results may also
be found in Ref. \onlinecite{braunbertram}.

\section{relation to the double sine-Gordon system}

In this section it is shown  that the
results of the previous sections allow us to
calculate the nucleation rate of kink-antikink pairs
of the double sine-Gordon model for
moderate to large friction.  Second, we shall see
that in the limit $Q^{-1} \delta^2\to \infty$
(i.e., large hard-axis anisotropy
and/or fields close to  the anisotropy field),
magnetization reversal rates
become equivalent to the creation rates of
kink-antikink pairs in the double sine-Gordon model.

Discarding noise terms for the moment,
we consider the dynamics of a field variable $\phi(x,t)$ which
is governed by the damped  double sine-Gordon equation
\begin{equation}
Q\partial_t^2 \phi +\alpha \partial_t \phi =- {\delta {\cal E}_{\rm dSG}
\over \delta \phi},
\label{dsg1}
\end{equation}
with the energy
\begin{equation}
 {\cal E}_{\rm dSG}=\int_{L/2}^{L/2}  dx \left\{ {1\over 2}
(\partial_x \phi)^2 + {1\over 2} \sin^2 \phi  - h \cos \phi
\right\}.
\label{dsg2}
\end{equation}
The constant $Q$ plays the role of  a mass and $\alpha$
is a damping constant. In the overdamped limit, the inertia
term $Q\partial_t^2 \phi$ in (\ref{dsg1})
can be neglected and the dynamics is purely determined
by the damping term.  Note that (\ref{dsg2}) is equivalent to
the energy density
(\ref{e}) restricted to the easy-plane $\theta=\pi/2$.
Therefore,  Eq. (\ref{dsg1}) exhibits the same
saddle point solution $\phi_s$ (\ref{nucl}) as
the full magnetic system. The corresponding barrier
energy  between the metastable state
$\phi=\pi$ and the absolute minimum
$\phi=0$ is given by ${\cal E}_s$ (\ref{nuclen}).
Expanding $\phi(x,t)=\phi_s(x) +\varphi(x,t)$
and  linearizing the equation of motion
(\ref{dsg1}) around the saddle point yields
\begin{equation}
Q\partial_t^2 \varphi +\alpha \partial_t\varphi +{\cal H}^{s\varphi}
\varphi=0,
\label{dsgfluct}
\end{equation}
where ${\cal H}^{s\varphi}$ is given by (\ref{hfop}).
A description of the stochastic dynamics in the vicinity
of the saddle is obtained by adding the stochastic force
$\xi(x,t)$ to the r.h.s. of (\ref{dsgfluct}).
with the noise correlation $\langle \xi(x',t') \xi(x,t) \rangle =
(2 \alpha/\tilde \beta)\delta(x-x')\delta(t-t')$.
The corresponding Fokker-Planck equation takes the
form of (\ref{FP}) if we identify $\tilde\beta=\beta {\cal A}$,
$M_{11}=0$, $M_{12}=-M_{21}=1$, $M_{22}=\alpha$
and ${\cal H}= ({\cal H}^{s\varphi}, Q^{-1})$.

The transition rate from the metastable state (neglecting
transitions that lead back from the metastable state
over the barrier to the initial state)
can then be calculated as in the appendix with the result:
\begin{equation}
\Gamma_{\rm dSG}= {\tilde\lambda_+ \over \sqrt{2 \pi^3} }
\sqrt{{\cal E}_s} L
\sqrt{\tilde \beta} \sqrt{ {\mathop{\rm det}\nolimits {\cal H}^{m\varphi}
\over \mathop{\rm det'}\nolimits |{\cal H}^{s\varphi}| }}
e^{-\tilde \beta {\cal E}_s},
\label{dsgrate}
\end{equation}
where ${\cal E}_s$ is given by  (\ref{nuclen}).
In (\ref{dsgrate}) a factor of 2 has been included
due to the existence of two
equivalent saddle points $\pm\phi_s$,
and the ratio of the determinants
has been calculated in (\ref{detfex}).
$\tilde\lambda_+>0$ is the nucleus decay
frequency which is obtained by insertion of
$\varphi =e^{\tilde\lambda_+ t}  \varphi_+$
into (\ref{dsgfluct}),
\begin{equation}
\tilde\lambda_+={\alpha Q^{-1} \over 2}
[-1 + \sqrt{1 + 4Q|E_0^{s\varphi}|/\alpha^2}],
\label{dsglambda}
\end{equation}
with $E_0^{s\varphi}$  the (negative) ground state
energy of ${\cal H}^{s\varphi}$. Eqns (\ref{dsgrate}), (\ref{dsglambda})
constitute the creation rate of kink-antikink pairs
in the moderately damped double sine-Gordon system.

It may now be verified that the magnetization reversal rate
(\ref{rate}), (\ref{pref}) with (\ref{pdet1}) is equivalent to the
result (\ref{dsgrate}) in the limit
$Q^{-1}\delta^2\to\infty$,  provided
that the time scale in (\ref{dsg1})-(\ref{dsgrate}) is chosen
as $[t]=M_0/(2\gamma K_e)$ while energies and lengths are
chosen as in (\ref{units}).  Taking the limit
$Q|E_0^{s\varphi}|/\alpha^2\to 0$  in
(\ref{lalargeR}) and (\ref{dsglambda}) we obtain $\lambda_+=
(\alpha+\alpha^{-1}) | E_0^{s\varphi}|$ and
$\tilde\lambda_+ =| E_0^{s\varphi}|/\alpha$, respectively.
Reinstating units, the equivalence of
$\Gamma$ and $\Gamma_{\rm dSG}$ is immediately verified.

\section{discussion}

In the previous sections we have investigated the rate of
magnetization reversal in an effectively 1D ferromagnet
which describes magnetization configurations in an
ideal elongated particle of a small constant cross section ${\cal A}$.
The experimentally most important conclusion is the existence of
a saddle point structure which is {\it localized} along the
sample.   Unlike the N\'eel-Brown theory \cite{neel} which leads to a
barrier energy $V K_e(1-h)^2$ proportional to the particle
volume $V$, the present theory leads to an energy barrier
${\cal A}{\cal E}_s$ that is proportional  to the sample cross section  and
to the domain wall energy (after reinstating the units (\ref{units})).
For sufficiently elongated particles the energy of the nonuniform
barrier is thus always lower than that of the uniform one and thus
the present theory predicts much lower coercivities than
the  N\'eel-Brown theory. To illustrate this, we consider
the following typical material parameters of particles
such as $CrO_2$: $A=5\cdot10^{-7}$ erg/cm,
$K_e=7\cdot10^5$, ${\rm erg/cm}^3$, $M_0=480$ Oe,
$\gamma=1.5\cdot10^7 {\rm Oe^{-1} s^{-1}}$.
For $T=300$ K, $Q^{-1}=0.2$, $\alpha=0.05$,
the numerically evaluated
switching rate (\ref{rate}), (\ref{omfexact})
is  shown in Fig. \ref{fig6} as a
function of the external field
for various particle diameters but for a fixed aspect ratio of
15:1.  The dotted lines represent the predictions \cite{KlGu} of
the N\'eel-Brown theory, while  $H_{\rm ext} M_0/(2 K_e)=1$
is the Stoner-Wohlfahrth value of the nucleation field.
One clearly recognizes the dramatic coercivity reduction
for particles with small diameters.  Note that the switching rate
at a given field predicted by the present theory exceeds that
of the N\'eel-Brown theory by more than 10 orders of magnitude.
Conversely, the coercivity of a particle of diameter $100 \AA$
exhibits a coercivity which (depending on the measurement time)
is about one third of the Stoner-Wohlfahrth value.
Since the barrier energy ${\cal A} {\cal  E}_s$ is independent of
the particle length, the present theory predicts the coercivity
to become independent of the particle length for sufficiently
long particles. This is in contrast to the theory of N\'eel and Brown
which predicts a suppression of thermal effects in the particle
volume.

Experiments investigating the coercivity of a single elongated particle
indeed show a significant coercivity reduction from
the Stoner-Wohlfahrth value for fields
along the particle.
These experiments have also shown an asymmetry in the
angular dependence of the coercivity,  the
coercivity reduction being more pronounced
for external fields along the particle axis
than for fields directed perpendicularly to the sample.
Both of these findings are in qualitative agreement \cite{braunbertram}
with the present theory.
A quantitative comparison between theory and experiments
is difficult for the presently available experimental data
since the particles are irregularly shaped and often contain voids.
Experiments on  particles with a more perfect morphology
such as $CrO_2$ or  data of particles with various aspect ratios
and diameters would clearly be desirable to further clarify
the mechanism of thermally activated magnetization reversal.

Let us now recall the various assumptions that have been
made in the present theory:

The cross-sectional area has been considered
constant throughout the particle.  This assumption leads
to  a continuous degeneracy of the solution  $\phi_s$
with respect to translations.  In the case of
a varying cross sectional area, the present treatment will still be
approximately correct if  the variations have a much shorter
wavelength than the characteristic length scale $\delta$ of the
nucleus. However, if the cross sectional area varies
substantially, the saddle point energy will depend
on the coordinate $x_0$ in (\ref{nucl})
and hence a whole class of energetically almost degenerate
saddle points  emerge. Such an extension of
the present theory would predict that one single particle
behaves as if there would be a distribution of saddle point energies.
Experimental results \cite{lederman} indeed show deviations
from an Arrhenius law involving  a single energy barrier.
They exhibit a decay of the magnetic moment of a single particle
 that is proportional to $\ln t$ over several decades in $t$, a
fact that is usually attributed to a distribution  of  energy barriers.
Such a behavior cannot be reconciled with the simple N\'eel-Brown
picture which predicts a unique energy barrier for a single particle.

In addition, we have focussed on nucleation in the interior of the particle
but we have neglected effects occurring at the particle ends:

Since the nucleus $\phi_s$ describes a magnetization configuration
merging asymptotically  into the metastable state, the present
nucleus may also be used to describe a situation
where the  magnetization is  pinned
at the sample ends.   In order for the present theory to hold, however,
the pinning energy has to be sufficiently small that it can be
overcome by the two domain walls propagating from the
nucleation location to the sample ends.

In the opposite case of free boundary conditions,
i.e. ${\bf M}'(\pm L/2)=0$, there exists also the possibility
that  only one domain wall is nucleated at one sample end.
This case can also  be related to the present theory.
In the ideal  situation of a sample of constant cross section
and  an effective easy-axis anisotropy that extends to the sample end
(at least within a distance smaller than the domain wall width),
the saddle point structure $\phi_s$ restricted to the interval
$-\infty<x<0$ represents a domain wall which is nucleated
at the sample end $x=0$. Consequently the
corresponding energy is half of the nucleus energy
${\cal A}{\cal E}_s$.

The theory as outlined in the appendix  applies to the
regime of moderate to large friction. Since, however,  the
damping constant in magnetic systems is quite small,
some estimates of the applicability range
of the present theory are presented in the following.

\subsection{Validity of the Theory}

The principal existence of a
lower limit of the damping constant for the present
theory may be seen as follows.  For $\alpha=0$, the
linearized equations (\ref{linLLG}) do not describe
the decay of the nucleus towards the stable state
 but rather a purely precessional motion which
conserves the energy. Therefore,  the corresponding decay frequency
$\lambda_+$ is completely irrelevant for the
nucleation rate for $\alpha\to 0$.

For very small values of the damping constant $\alpha$, a completely
different methodology would have  to be applied, since the
nucleation does no longer correspond to a diffusion in
configuration space but rather in energy space.
Since the  time evolution of the nucleus for extremely small $\alpha$
is  expected to exhibit  a ``breathing" oscillation \cite{braunbrod},
a derivation of the corresponding  Fokker-Planck equation would be an
extremly difficult task. However, we shall see that the
applicability range of the  present theory extends
to rather small values of $\alpha$ even for small
nuclei and small cross sectional areas  of the sample. Therefore
we do not consider the underdamped theory any further.

A criterion for the crossover between  the present theory and
energy diffusion has been given by Landauer and Swanson
(Ref.\onlinecite{Brink} and in Ref. \onlinecite{hanggi}
(see also Ref. \onlinecite{HBB})).
The moderately damped theory may be applied  if the energy loss during
an  (approximate)  period of the motion near the saddle point
exceeds $k_B T$.

Using the equations of motion (\ref{LLG0}) we obtain for the
energy loss rate per area
\begin{equation}
{d{\cal E}\over d t}=-\alpha \int dx
\left(\partial_t{\bf m}\right)^2,
\label{enloss}
\end{equation}
where ${\bf m}\equiv {\bf M}/M_0$.
Since (\ref{enloss}) may also be expressed by spatial instead of temporal
derivatives it is clear that the energy loss will be smallest for small nuclei.
In this limit we may employ the spin wave approximation
$m_y,m_z\ll 1$, $m_x=-1+{\cal O}(m_x^2)$. The energy loss
during one approximate period then takes the following form
\begin{equation}
{\cal A} \Delta {\cal E}=-\alpha {\cal A}\int_0^{2\pi\over\omega} dt
\int dx \left\{(\partial_t m_y)^2 + (\partial_t m_z)^2\right\}
\label{periodloss}
\end{equation}
where $2\pi\over \omega$ is the precession period.
The r.h.s. of (\ref{periodloss}) is now evaluated approximately.
First, we are only interested in leading order in $\alpha$ and thus we
may use the conservative equations of motion for the evaluation
of the integrand in (\ref{periodloss}). Second we discard any
breathing effects and
neglect the exchange coupling of the magnetic moments
such that the precession amplitude is given by the
spatial distribution of the nucleus.
Linearization of the equations of motion for $\alpha=0$,
$\partial_t {\bf m}={\bf m} \times
(\partial {\cal E}_0/\partial {\bf m})$, with ${\cal E}_0$  the
energy (\ref{eb1}) without exchange,
then leads to
\begin{eqnarray}
\partial_t m_y&=&(1-h+Q^{-1}) m_z \nonumber\\
\partial_t m_z&=&-(1-h)m_y.
\label{mlinear}
\end{eqnarray}
They describe an elliptical precession
$m_z=m_z^0 \sin\omega t$,
$m_y=-m_y^0\cos\omega t$
with $m_z^0/m_y^0=$ $\sqrt{1-h}/\sqrt{1-h+Q^{-1}}$ and
$\omega=\sqrt{(1-h)(1-h+Q^{-1})}$.
Inserting this into (\ref{periodloss}) we obtain
\begin{equation}
{\cal A}\Delta {\cal E}=-\alpha {\cal A} {2\pi \over \omega}
\int dx (m_y^0)^2 (1-h) (1-h+{Q^{-1}\over 2}).
\label{totalloss}
\end{equation}
Now, for small nuclei, we have $h=\mathop{\rm sech}
\nolimits^2R\simeq 1-R^2$. The precession
is assumed to cover the nucleus structure and therefore we have
$m_y^0=2\sinh R\cosh(x/\delta)/(\sinh^2 R+\cosh^2(x/\delta))$.
Inserting this in (\ref{totalloss}) and performing the integration
the validity condition ${\cal A} \Delta {\cal E}> k_B T$ takes the
following form in leading order $R$
\begin{equation}
\Delta {\cal E} = 8 \pi Q^{-1/z2} \beta {\cal A} \alpha R^2>1.
\label{limit}
\end{equation}
For typical values of the constants as given above  and $Q^{-1}=0.2$
we obtain the condition $\alpha R^2 a\;7\cdot 10^{13}>1$
where $a$ is the cross section area measured in ${\rm cm}^2$.
The limit is reached for e.g.  $\alpha=0.05$,
$R=0.4$ ($h=0.84$) at particle diameters  $70 \AA$.
Note that due to the neglection of the
breathing contribution to the energy loss this represents
a lower limit for $\alpha$ and therefore the theory may be applied even for
smaller values of $\alpha$ or samples smaller than indicated above.

\section{acknowledgements}

I kindly acknowledge illuminating
discussions with W. Baltensperger, S. Skourtis,
H. Suhl and P. Talkner.  This work was supported
by the Swiss National Science Foundation
and  by ONR-Grant  N00014-90-J-1202.

\appendix
\section*{}

In this appendix we present a derivation of the expression for the
nucleation rate
starting from the Fokker-Planck equation (\ref{FP}) near the saddle point.
Since the deterministic part of the dissipative linearized equations
(\ref{lambda}) constitutes a non Hermitian
eigenvalue problem, there is no known set of
eigenfunctions to expand  in and therefore traditional formulations
which are based on an expansion into mode amplitudes
cannot be applied immediately.  In the sequel we shall show,
however, that
the within a functional formulation, the derivation of the nucleation
rate can be carried out
in close analogy to the methods of Kramers \cite{Kramers} and
its extension to many degrees of freedom  by Langer \cite{Langer}$^{(b)}$.
In the limit of large friction the result reduces to the
theories\cite{Brink} of Brinkman, Landauer-Swanson and Langer.

The basic principles of the method go back to  Kramers \cite{Kramers}.
We calculate the stationary flux of a nonequilibrium distribution
across a surface transverse to the unstable direction at the saddle point.
A main difference to finite dimensional problems
is the existence of the Goldstone mode with zero energy
which reflects the continuous degeneracy of the nucleus $\phi_s$
with respect to rigid translations.

Our goal is the construction of  a {\it stationary nonequilibrium}
probability density obeying the boundary conditions
$\varrho\simeq 1$ near the metastable
state and $\varrho\simeq 0$ beyond the saddle point.
To this end we factorize the desired probability distribution
as follows:
\begin{equation}
\varrho = \varrho_{\rm eq} F.
\label{fact}
\end{equation}
The key assumption is now to let $F$ be a function of one coordinate $u$
only. $F$ is such that $\varrho$ is a normalizable function. In the vicinity
of the saddle point, the coordinate $u$ is a linear functional
in $\psi$
\begin{equation}
u=\int dx \sum_j U_j(x)\psi_j(x),
\label{u}
\end{equation}
with $(U_1,U_2)=(U^{\varphi},U^p)$.
After insertion of  (\ref{fact}) with (\ref{u}) into (\ref{FP}) and using
the stationarity of $\varrho$ we obtain
\begin{eqnarray}
\int &dx&\sum_{ij}\;M_{ij} \left[-{\cal H}_i \psi_i(x) U_j(x) {dF\over du}
\right.\nonumber\\
&&\left.+{1\over \beta{\cal A}} U_i(x) U_j(x) {d^2F\over du^2}\right]=0.
\label{Fu}
\end{eqnarray}
In order for (\ref{Fu}) to become a proper differential equation in
$u$ alone we must have
\begin{equation}
\int dx \sum_{ij}\,M_{ij}{\cal H}_i \psi_i (x) U_j(x) =\kappa u,
\label{kappa}
\end{equation}
\begin{equation}
{1\over \beta {\cal A}}\int dx \sum_{ij}M_{ij} U_i(x) U_j(x) = \gamma \kappa,
\label{gkappa}
\end{equation}
where $\gamma$, $\kappa$ are real constants.
Since from (\ref{M}),  $\sum_{ij}M_{ij} U_i U_j =\alpha\sum_i U_i ^2$,
the condition (\ref{gkappa}) amounts to a normalization of the $U_i$
and we have $\gamma\kappa>0$.
Using (\ref{kappa}) and (\ref{gkappa}), Eq (\ref{Fu}) reduces to the
same differential equation as in the
case\cite{Kramers} of one degree of freedom:
\begin{equation}
-u{dF\over du}+\gamma{d^2 F\over du^2} =0.
\label{Fuconst}
\end{equation}
 which is integrated with the boundary conditions $F(-\infty)=1$,
$F(\infty)=0$,
i.e.  the vicinity of  the metastable state  ($u\to -\infty$)
is characterized by thermal equilibrium while the probability distribution
vanishes beyond the sadlle point ($u\to \infty$).
Since $\varrho$ has to be normalizable we must have
$\gamma<0$ and hence $\kappa<0$. $F$ is then given
by
\begin{equation}
F(u)={1\over \sqrt{2\pi|\gamma|}}\int_u^ \infty du \exp
\left\{-{u^2\over 2|\gamma|} \right\}.
\label{F}
\end{equation}
Inserting  (\ref{F}) and (\ref{gkappa}) into (\ref{current}) we
obtain for the current near the saddle point:
\begin{equation}
J_i(x)={1\over \beta{\cal A}\sqrt{2\pi|\gamma|}}\exp
\left\{-{u^2\over 2|\gamma|} \right\} \varrho_{\rm eq}
\sum_j M_{ij} U_j(x).
\label{Ji}
\end{equation}
\bigskip
We now return to the evaluation of $\kappa$.
Equation (\ref{kappa}) is fulfilled if
\begin{equation}
\sum_j M_{ij} {\cal H}_i U_j =\kappa U_i.
\label{kappadef}
\end{equation}
Note that this implies that $U^{\varphi}(x)$ has no component along
the zero frequency mode since $(\chi^{s\varphi}_1,U^{\varphi})=$
$\kappa^{-1}$ $\cdot\sum M_{\varphi j}$ $\cdot({\cal H}^{s\varphi}
\chi^{s\varphi}_1,U^{\varphi})=0$. This is physically
plausible since the dynamical instability of $\phi_s$ is only associated
with a
shrinking or expansion of the nucleus but {\it not} with a pure translation of
the nucleus as described by $\chi^{s\varphi}_1(x)$.
Because ${\cal H}^{s p}$ is a positive operator there is no
restriction on the functions $U^p (x)$. Therefore (\ref{kappadef})
can be inverted to give
\begin{equation}
\sum_{j}M_{ij} U_j =\kappa {\cal H}^{-1}_i U_i,
\label{inv}
\end{equation}
where $({\cal H}^{s\varphi})^{-1}$ acts only on the
subspace $\{f|(f,\chi^{s\varphi}_1)=0\}$. Eqn (\ref{inv}) and hence
(\ref{kappadef}) are solved by putting
$U_i={\cal H}_i\psi_i^+$ where  $\psi^+=(\varphi_+,p_+)$ is the dynamical
unstable mode obeying (\ref{lambda}). Therefore we have
\begin{equation}
\kappa=-\lambda_+<0.
\end{equation}
The escape rate $\Gamma$ is now obtained by
integrating the flux transverse to the unstable
direction, e.g. over the manifold $u=0$:
\begin{eqnarray}
\Gamma&=&\int_{u=0}{\cal D}\varphi{\cal D} p \int dx
\sum_i U_i(x)J_i(x)\nonumber\\
&=&\lambda_+
\sqrt{{|\gamma|\over2\pi}} \int \!{\cal D}\varphi{\cal D} p \,
\delta(u) \,\varrho_{\rm eq} .
\label{gamma}
\end{eqnarray}
where we have used (\ref{Ji}) and $\lambda_+$ is determined
by the eigenvalue equation (\ref{lambda}).
The constant $\gamma$ will cancel in
the final result as we shall see below. $\varrho_{\rm eq}$
has to be evaluated on the hyperplane $u=0$ in the
vicinity of the saddle point. According to (\ref{req}) it is given  by
$\varrho_{\rm eq}=\exp\{-\beta{\cal A}{\cal E}_s^{(2)}\}/Z$ with . $Z$ is
determined by the normalization of $\varrho_{\rm eq}$ in the
vicinity of the metastable state.  Upon  insertion of (\ref{req}) and
using the Fourier representation
of the $\delta$-function, eq.  (\ref{gamma}) reads
 \begin{eqnarray}
\Gamma&=&\lambda_+\sqrt{|\gamma|\over
2 \pi} {e^{-\beta{\cal A}{\cal E}_s}\over Z}\int {dq\over 2\pi}
\int {\cal D}\varphi\,{\cal D} p\; \exp\left\{ iqu\right\}\nonumber\\
 &\times&\exp\left\{-\beta{\cal A}[{1\over 2}\int dx
\varphi{\cal H}^{s\varphi}\varphi+{1\over 2}
\int dx p{\cal H}^{s p} p]\right\} .
\label{G1}
\end{eqnarray}
For the evaluation of the functional integrals
we use the self adjointness of ${\cal H}^{s\varphi}$ and ${\cal H}^{s p}$
acting
on functions with periodic boundary conditions on
$[-{L \over 2},{L\over 2}]$. Therefore they have
an orthogonal and complete set of eigenfunctions
$\chi^{s\varphi}_\nu$,$\chi^{sp}_\nu$ with ${\cal H}^{s\varphi}
\chi^{s\varphi}_\nu=E^{s\varphi}_\nu \chi^{s\varphi}_\nu$,
${\cal H}^{s p}\chi^{sp}_\mu=E^{sp}_\mu \chi^{sp}_\mu$, where $\nu,\mu$
denote the bound states $\nu, \mu=0,1,(2)$ and  scattering states
$\nu, \mu=k$  as well.
Therefore we can expand
\begin{eqnarray}
\varphi(x)&=&\sum_\nu\varphi_\nu\,\chi^{s\varphi}_\nu(x),\label{fexpan}\\
p(x)&=&\sum_\mu p_\mu\,\chi^{sp}_\mu(x),\label{pexpan}
\end{eqnarray}
where $\chi_{-k}=\chi_k^*$, and $\varphi_\nu, p_\nu$ are
complex expansion coefficients.
Since $\varphi$ and $p$ are real we have
$\varphi_\nu=\varphi_{-\nu}^\ast$ and
$p_\nu=p_{-\nu}^\ast$.
Expanding similarly $u=\sum U^{\varphi\;\ast}_\nu
\varphi_\nu +\sum U^{p\;\ast}_
\mu p_\mu$
we obtain
\begin{eqnarray}
\Gamma&=&\lambda_+\sqrt{|\gamma|\over 2 \pi}
{e^{-\beta{\cal A}{\cal E}_s}\over Z}
\int\! {dq\over 2\pi}
\int\!\prod_\nu d\varphi_\nu\int\! \prod_\mu dp_\mu\nonumber\\
&\times&\exp\left\{ \sum'_\nu (iqU^{\varphi\;\ast}_\nu
\varphi_\nu -
{\beta{\cal A} \over 2}E^{s\varphi}_\nu\varphi_\nu^{\;\ast}
\varphi_\nu)\right\}
\nonumber\\
&\times&\exp\left \{ \sum_\mu(iqU^{p\;\ast}_\mu p_\mu -
{\beta{\cal A} \over 2}E^{sp}_\mu p_\mu^{\;\ast}  p_\mu)\right\},
\label{G2}
\end{eqnarray}
where the prime on the sum indicates that
the integrand is independent of $\varphi_1$, since
$E^{s\varphi}_1=U^\varphi_1=0$.  To simplify notation
we now choose integration measures $(d\varphi_\nu)$ and $(d p_\mu)$
in (\ref{G2}) {\it and} in $Z$ such that e.g.
\begin{eqnarray}
\int_{-\infty}^{\infty} (dp_0) \exp\left\{-{\beta{\cal A}\over 2}E^{sp}_0
p_0^2\right\}={1\over\sqrt{E^{sp}_0}},\nonumber\\
 \int_{-\infty}^{\infty}
(dp_k^\ast)(dp_k)\exp\{-\beta{\cal A}E^{sp}_k p_k^\ast p_k\}=
{1\over E^{sp}_k}.
\label{measure}
\end{eqnarray}
But note that we have to restore $\sqrt{\beta{\cal A}/2\pi}$
in the integration over the zero
mode which does not contain a Gaussian and note also
that the integration measure in $q$ is the usual
one. We now perform the integrations in
$\varphi_\nu$ and $p_\mu$ except for the amplitudes
of the zero mode $\varphi_1$ and that of the unstable mode,
$\varphi_0$, to obtain:
\begin{eqnarray}
\Gamma&=&\lambda_+\sqrt{|\gamma|\over 2 \pi}
{e^{-\beta{\cal A}{\cal E}_s}\over Z}
{1\over \sqrt{\det''{\cal H}^{s\varphi}}}
{1\over \sqrt{\det{\cal H}^{s p}}}\int (d\varphi_1)\nonumber\\
&\times&\int \!(\!d\varphi_0\!) \int {\!dq\over 2\pi}
\exp\left\{\!-{q^2\over 2\beta{\cal A}}
(\sum\nolimits_\nu^{''}{|U^\varphi_\nu|^2\over E^{s\varphi}_\nu}+
\sum\nolimits_{\mu}{|U^p_\mu|^2\over E^{sp}_\mu})\!\right\}\nonumber\\
&\exp&\left\{iqU^{\varphi}_0\varphi_0 + {\beta{\cal A}\over 2}
|E^{s\varphi}_0|\varphi_0^2 \right\}.
\label{G3}
\end{eqnarray}
The expression $\det{\cal H}$ stands for the product of all
eigenvalues of ${\cal H}$. The double prime on the determinant
and the summation indicates that the terms $\nu=0,1$ corresponding to
the unstable mode and the zero mode are omitted.

We now turn  to the evaluation of the zero mode.
To this end we remember that a pure
translation of the nucleus can be described by
$\phi_s(x+dx)-\phi_s(x)=\phi_s'(x) dx$ or since
$\chi^{s\varphi}_1\propto d\phi_s/dx$ we can equally well
write $\phi_s(x+dx)-\phi_s(x)=\chi^{s\varphi}_1(x)\,d\varphi_1$.
This allows us to replace the integration over the
zero mode by an integration over $x$.
Since $\chi^{s\varphi}_1$ is normalized to
unity and reinstating the integration measure (\ref{measure}) of
$(d\varphi_1)$ we have
\begin{equation}
\int (d\varphi_1)=\sqrt{{\beta{\cal A}\over2\pi}}\int_{-L/2}^{L/2} dx
\sqrt{\int dx'
\left[ {d\phi_s\over dx'}\right] ^2}=\sqrt{{\beta{\cal A}\over2\pi}}{\cal L},
\label{zeroint}
\end{equation}
where
\begin{equation}
{\cal L}\equiv\sqrt{{\cal E}_s} L.
\label{Ldef}
\end{equation}
Here we have made use of (\ref{nuclen}). After performing the
$q$- integration (\ref{G3})  we have to convince ourselves that
the remaining integral converges. This is ensured by the following
relation, which follows from (\ref{gkappa}) and (\ref{inv})
\begin{equation}
\sum_i\int dx U_i(x) {\cal H}_i^{-1}U_i(x) = -|\gamma| \beta{\cal A}<0.
\label{negexp}
\end{equation}
Using the expansions of $U_i$ in terms of the
$\chi_\nu^{s\,\varphi,p}$ and recalling that
$U_1^\varphi\equiv 0$ we see that the exponent is in fact negative.
We are now left with the evaluation of $Z$.
Since $\varrho_{\rm eq}$ is strongly peaked at the
metastable state we can perform a Gaussian approximation
\begin{eqnarray}
Z&=&\int {\cal D}\varphi\,{\cal D} p \exp
\left\{-\beta{\cal A}[ {1\over2}\int dx\;
\varphi\;{\cal H}^{m\varphi}\;\varphi\right.\nonumber\\
&&\left. +{1\over 2}
\int dx\; p\;{\cal H}^{m p}\;p]\right\},
\label{Zm}
\end{eqnarray}
where in contrast to the above $\varphi$ and $p$ now
describe fluctuations out of the {\it metastable}
state. They are defined by $\phi=\pi +\varphi$ and
$\theta=\pi/2-p$, where $|\varphi|,|p|\ll 1$. This
integral is performed analogously to the previous one:
$\varphi$ and $p$ may be expanded into
the (plane wave) eigenstates of ${\cal H}^{m\varphi},{\cal H}^{m p}$.
The integrations then are all Gaussian and
using the measure (\ref{measure}) for the integrations, we finally obtain
\begin{equation}
Z={1\over \sqrt{\det{\cal H}^{m\varphi}}}{1\over \sqrt{\det{\cal H}^{m p}}}.
\end{equation}
After having performed the $q$-integration we can now
carry out the final Gaussian integration  over the unstable mode
$\varphi_0$.  Using  (\ref{gkappa}), (\ref{G3}) leads to  the final result:
\begin{equation}
\Gamma={\lambda_+\over 2\pi} {\cal L}  \sqrt{{\beta{\cal A}\over 2\pi}}
\sqrt{{\det{\cal H}^{m\varphi}\over
\det'|{\cal H}^{s\varphi}|}}\sqrt{{\det{\cal H}^{m p}\over
\det{\cal H}^{s p}}}e^{-\beta{\cal A}{\cal E}_s},
\label{Gres}
\end{equation}
where $\det'|{\cal H}^{s\varphi}|$ denotes the product of the modulus
of the eigenvalues with omission of the zero mode.
 Apart from the dynamical prefactor the result {\it looks as if }
we had evaluated in Gaussian approximation the
ratio $Z_m/Z_s$ of the partition functions at the metastable
state and the saddle point, respectively,  with the unstable
mode rendered to a stable one.  Or loosely speaking, the
nucleation rate is proportional to the imaginary part of
the ratio $Z_m/Z_s$. For later reference, we  express
 the result (\ref{Gres})  in terms of the eigenvalues
$E^{s\varphi}_\nu$,$E^{sp}_\mu$:
\begin{eqnarray}
\Gamma&=&{\lambda_+\over 2\pi}{\cal L}  \sqrt{{\beta{\cal A}\over 2\pi}}
{1\over\sqrt{|E^{s\varphi}_0|
E^{s\varphi}_2 E^{sp}_0 E^{sp}_1}}\nonumber\\
&\times&
\sqrt{{\prod_{k_1} E^{m\varphi}_{k_1}  \over \prod_{k_1'}
E^{s\varphi}_{k_1'} }}
\sqrt{{\prod_{k_2} E^{mp}_{k_2}  \over \prod_{k_2'}E^{sp}_{k_2'} }}
e^{-\beta{\cal A}{\cal E}_s},
\label{Gres2}
 \end{eqnarray}
where according to (\ref{Ldef}) ${\cal L}=L\sqrt{{\cal E}_s}$
(${\cal E}_s$ is given by (\ref{nuclen}) )
and $\lambda_+$ is
determined by (\ref{lambda}). Note that in our
case of two equivalent saddle points $\phi^\pm$ we have to
multiply this final result by a factor of two.

\begin{figure}
\caption{Various  anisotropy configurations axes that can be
described by the energy density
(\protect{\ref{e}}). The sample cross sectional areas
are assumed to be sufficiently small  such that
transversal variations in the magnetization
are suppressed.}
\label{fig0}
\end{figure}

\begin{figure}
\caption{The spatial variation of the
nucleus is shown for i) fields close to the
anisotropy field and ii) small fields.
In these pictures, the
magnetic chain is taken along the easy-axis.
However, the model (\protect{\ref{e}})
equivalently applies to all situations shown in
Fig \protect{\ref{fig0}}.
}
\label{fig1}
\end{figure}

\begin{figure}
\caption{
The reduced prefactor in the overdamped
limit is shown as a function
of the parameter $R$ for different values
of the hard-axis anisotropy. The lines
(-$\cdot$-$\cdot$) and (- - -) are the
asymptotic formulas (\protect{\ref{omsmallR}}) and
(\protect{\ref{omlargeR}}) respectively.}
\label{fig3}
\end{figure}

\begin{figure}
\caption{
The decay frequency of the nucleus is
shown for different values of $R$ and
$\alpha$. The dots are results of a numerical
solution of (\protect{\ref{lambda}}) and the solid
line is the approximation formula
(\protect{\ref{lalargeR}}).}
\label{fig4}
\end{figure}

\begin{figure}
\caption{
Numerical results  for the reduced prefactor
(\protect{\ref{omfexact}}) for moderate damping
as a function of $R$. The dashed lines
represent the asymptotic formula
(\protect{\ref{preflargeR}}).}
\label{fig5}
\end{figure}

\begin{figure}
\caption{
The total nucleation rate $\Gamma$ is shown
as a function of the reduced external field
for various particle diameters. The  material parameters
are chosen as in Sec. VIII. The dashed
curves indicate the results of the N\'eel-Brown theory. The particle
aspect ratio is assumed  to be 15:1.}
\label{fig6}
\end{figure}

\clearpage
\widetext
\begin{table}
\caption{Summary of the results for the prefactor $\Omega$
(\protect{\ref{omfexact}})
in the moderately damped limit
for i) large effective hard-axis anisotropy $Q^{-1}\delta \gg 1$,
($\delta=\coth R$),
and ii) small fields $R\gg 1$ ($h={\rm sech}^2 R$).
Note that the underlying dimensionless units are given by
(\protect{\ref{units}}).}

\begin{tabular}{l|c|c|c|}
 & $\Omega/L\sqrt{\beta{\cal A}}$
    &$\lambda_+$
        & $E_0^{s\varphi}$ \\ \tableline
%
$i)\;\; Q^{-1} \delta^2 \gg 1$\tablenote
         { In this limit, ${\rm det} {\cal H}^{mp}/{\rm det}
{\cal H}^{sp}=1$, cf.
          (\protect{\ref{pdet1}})  }
			 &
        &$-{\alpha\over 2}[Q^{-1}+E_0^{s\varphi}] +$
            &$E_0^{s\varphi}$\tablenote{has to be evaluated numerically}\\
%
    &${4\lambda_+\over \pi^{3/2}}
    \tanh ^{3/2} R \sinh R$
        &\multicolumn{1}{c|}{
         +$\sqrt{   \left({\alpha\over 2}\right)^2
        [Q^{-2}-2Q^{-1}E_0^{s\varphi}] -Q^{-1}E_0^{s\varphi}  }$
          }
            &\\     \cline{1-1} \cline{4-4}
%
%
$Q^{-1} \delta^2 \gg 1$\tablenotemark[1],$R \ll 1$
    &\multicolumn{1}{r|}{ (\protect{\ref{smallRex}}) }
        &\multicolumn{1}{r|}{ (\protect{\ref{lalargeR}})\tablenote{
           Compared to (\protect{\ref{lalargeR}}), small terms of
           the order ${\cal O} ((QE_0^{s\varphi})^2)$ have been dropped.}}
             &\\ \cline{1-3}
%
$Q^{-1} \delta^2 \gg 1$\tablenotemark[1],$R \ll 1,$
    & ${4\sqrt{3}\over \pi^{3/2}}\sqrt{Q^{-1}} R^{7/2}$
        & $\sqrt{3 Q^{-1}} R$
             &$-3 R^2$ \\
$\sqrt{Q} R/\alpha\gg 1$\tablenote{typically realized in experiments}
    &\multicolumn{1}{r|}{(\protect{\ref{prefalR}})}
        &\multicolumn{1}{r|}{(\protect{\ref{prefsmallR}})}
             &(\protect{\ref{lasmallR}})\\ \cline{1-3}
%
$Q^{-1} \delta^2 \gg 1$\tablenotemark[1],$R \ll 1,$
    &${12\over \pi^{3/2}}\left(\alpha+\alpha^{-1}\right)
      R^{9/2}$
        & $3 \left(\alpha+\alpha^{-1}\right) R^2$
             &\\
$\sqrt{Q} R/\alpha\ll 1$\tablenote{For $\alpha\to\infty$, these results
merge into (\protect{\ref{omsmallR}}), (\protect{\ref{omlargeR}})
obtained in the overdamped limit.}
    &\multicolumn{1}{r|}{(\protect{\ref{prefRla}})}
        &\multicolumn{1}{r|}{(\protect{\ref{prefsmallR}})}
             &\\ \hline \hline
%
$ii)\;\; R \gg 1$
    &${4\lambda_+\over \pi^{3/2}} \tanh^{3/2} R \sinh R $
        &$-{\alpha\over 2}[Q^{-1}+E_0^{s\varphi}]+$
             &$-8 e^{-2R}$\\
    &$\times[\sqrt{Q} + \sqrt{1+Q}]^2$
         & $ +\sqrt{ \left({\alpha\over 2}  \right)^2
           [Q^{-2}-2Q^{-1}E_0^{s\varphi}]-Q^{-1}E_0^{s\varphi}  }$
             &(\protect{\ref{hatbound}}) \\
    &\multicolumn{1}{r|}{(\protect{\ref{preflargeR}})}
       &\multicolumn{1}{r|}{(\protect{\ref{lalargeR}})}
           &\\ \cline{1-3}
$R \gg 1,$
    &${16 \over \pi^{3/2}} (\alpha + \alpha^{-1})
     e^{-R}$
        &$8(\alpha + \alpha^{-1}) e^{-2R}$
             &\\
$\sqrt{Q} e^{-R}/\alpha \ll 1$\tablenotemark[5]
   &$\times [\sqrt{Q} + \sqrt{1+Q}]^2$
        &
             &\\
    &\multicolumn{1}{r|}{(\protect{\ref{preflargeR}})}
        &\multicolumn{1}{r|}{(\protect{\ref{lalargeR}})\tablenotemark[4]}
             &\\
 \end{tabular}
 \label{table1}
 \end{table}

\begin {references}
\bibitem{curling}
Note,  however,  that the existence of a saddle point of
curling symmetry does not ensure the possibility
of magnetization reversal within a continuum model.
For topological reasons due to the ``backbone" at the
cylinder axis, the magnetization along an infinite cylinder
cannot be reversed  via configurations that
are always tangential to the sample surface.
While this argument  prohibits the existence of
``curling"-nuclei for small particles with diameter of less than
a domain wall width, the discreteness of the lattice
reduces the above topological barrier
for large sample diameters such that saddle points of
``curling symmetry",  which are localized along the
particle, are indeed  expected to be
the relevant energy barriers.
\bibitem{threshold} Note that Ref. \onlinecite{barton} deals with zero
boundary conditions, i.e. $\chi(-L/2)=\chi(L/2)=0$ which leads
to a shift of  the spectrum of even eigenfunctions by $\Delta k=\pi/L$
compared to periodic boundary conditions.
The spectrum of odd parity eigenfunctions is identical for
periodic and zero boundary conditions.
\bibitem{zeroenergy}  A zero energy resonance occurs if the
Jost function for even or odd parity is zero for $k=0$.
\bibitem{detcomment}
The functions  $\chi^{s\varphi}_1$,  $\xi^{s\varphi}_1$
should be replaced by $\chi^{sp}_0$,  $\xi^{sp}_0$ respectively.
However,   $\chi^{sp}_0$ given by (\ref{zeromode})
is now a symmetric function
while $\xi^{sp}_0$ an antisymmetric function.
\bibitem{underdamped}
This statement is  true in
the moderate to strong damping
regime which is considered in the present paper.
We  do not consider  here the underdamped regime
which would require a completely different approach
by constructing a Fokker-Planck equation in the
energy variable. For a criterion separating these two
regimes see also section VIII.
\end{references}
\end{document}